\documentclass[aps, prb, reprint, amsmath, amssymb,
floatfix]{revtex4-2}
\usepackage{graphicx}
\usepackage{bm}			
\usepackage{tabularx}
\usepackage{xcolor}
\usepackage[colorlinks, citecolor=blue, linkcolor=blue]{hyperref}

\begin{document}
\title{Self-induced Floquet states via three-wave processes in synthetic antiferromagnets}
\author{Thibaut Devolder} 
\email{thibaut.devolder@cnrs.fr}
\author{Joo-Von Kim} 
\affiliation{Universit\'e Paris-Saclay, CNRS, Centre de Nanosciences et de Nanotechnologies, Palaiseau, France}
\date{\today}                                           
                       
%
%
\begin{abstract}
We present a mechanism for self-induced Floquet states involving acoustic and optical modes in synthetic antiferromagnets. By driving optical modes off-resonantly with radiofrequency fields in the canted antiferromagnetic state, limit cycles arising from the predator-prey dynamics of the acoustic and optical mode populations can appear. The cyclic growth and decay of these mode populations induce a time-periodic modulation of the canted state, which subsequently generates Floquet states. These states appear as a rich  frequency comb in the power spectrum of magnetization oscillations.
\end{abstract}

\maketitle

%
%


Time-periodic driving in many-body systems offers a powerful tool to tailor material properties dynamically. The emergent Floquet states, the temporal analogue of Bloch states arising in a spatial crystal, can possess modified band gaps or topologically nontrivial features not present at equilibrium~\cite{oka_floquet_2019, harper_topology_2020, rudner_band_2020, rodriguez-vega_low-frequency_2021}. Within the context of magnetic systems, Floquet engineering of magnons has been explored in quantum magnets where synthetic chiral interactions can be induced with circularly-polarised electric fields~\cite{owerre_floquet_2017}, in cavity magnonics~\cite{xu_floquet_2020, proskurin_mode_2022} where ultrastrong coupling to microwave photons can modify the level spacing between modes, in coupled magnonic waveguides where modulation controls exceptional points~\cite{wang_floquet_2023} and magnon-magnon coupling~\cite{li_floquet_2023}, and in hybrid superconducting systems where magnonic NOON states can be generated~\cite{qi_floquet_2023}.

Recently, it has been shown that time-periodic modulation of spin textures provides an efficient means to generate magnon Floquet states, such as eigenmodes about a gyrating vortex in thin-film disks~\cite{heins_self-induced_2026}. In contrast to usual approaches in Floquet engineering, where direct modulation of material properties are sought, the emergence of Floquet magnons within a gyrating vortex stems from the time-periodic modulation of the background magnetization on which the eigenmodes are defined. This texture-driven Floquet mechanism allows the ultrastrong coupling limit to be reached with driving magnetic fields of only a few milliteslas in amplitude. Moreover, it was demonstrated that nonlinear interactions between the magnon modes and the vortex dynamics can enable “self-induced” Floquet states. This occurs when large amplitude driving of bare magnon modes leads to self-sustained gyration of the vortex core, which in turn drives the appearance of Floquet states. In this light, it is interesting to enquire whether self-induced mechanisms could exist within other magnetic configurations.

Here, we discuss a comparatively simpler magnetic system in which self-induced Floquet states can form. It involves a synthetic antiferromagnet (SAF), comprising two ferromagnetic layers coupled together antiferromagnetically by an interlayer interaction of the RKKY form. Under high power field pumping, nonlinear magnon processes such as three-magnon splitting (3MS) can take place~\cite{boardman_three-_1988, mathieu_brillouin_2003, liu_time-resolved_2019, schultheiss_direct_2009, ordonez-romero_three-magnon_2009, schultheiss_excitation_2019, korber_nonlocal_2020, verba_theory_2021}, whereby pumped optical modes split into half-frequency acoustic modes~\cite{sud_electrically_2025, mouhoub_spin_2026}. For small frequency detunings, we observe predator-prey dynamics involving the optical and acoustic mode populations, corresponding to the periodic growth and decay of these populations. This in turn results in a periodic modulation of the canted magnetization ground state, thereby generating Floquet states. This route to mode population dynamics differs fundamentally from similar instabilities observed in saturated ferromagnetic systems resulting from four-magnon scattering~\cite{rezende_model_1986, rezende_self-oscillations_1992}, which is restricted to the MHz-scale self oscillations. Here, the lifetimes of the two highly-populated SAF modes are sufficiently long to experience the GHz-scale periodic modulation of the ground state of the system on which they reside, thereby leading to Floquet bands with a double frequency comb as spectral signature.

\begin{figure}
\centering\includegraphics[width=8.5cm]{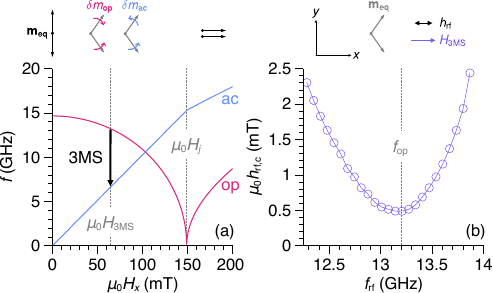}
\caption{(a) Dependence of the eigenmode frequencies on the applied static field, $H_x$, with the sketch of the equilibrium magnetizations of the SAF shown in the top insetl. The  3MS arrow indicates the matching condition, i.e., where $\omega_\mathrm{op} = 2\omega_\mathrm{ac}$. (b) Threshold pumping field for 3MS, $h_{\mathrm{rf},c}$, as a function of frequency detuning between the pumping frequency $f_\mathrm{rf}$ and $f_\mathrm{op}$ at $H_x = H_\mathrm{3MS}$. The top inset illustrates the field configuration.}\label{GeneralDescription}
\end{figure}

We consider a model SAF comprising two identical, uniformly-magnetized layers $\ell=1, 2$, which are coupled together by an interlayer exchange interaction favoring antiparallel alignment [Fig.~\ref{GeneralDescription}(a)]. We assume material parameters drawn from recent experiments on CoFeB/Ru/CoFeB SAFs~\cite{adam_role_2025, mouhoub_spin_2026}, i.e., a saturation magnetization of $M_s=1.35$~MA/m, a Gilbert damping constant of $\alpha=0.011$, and an interlayer exchange field of $H_j = 119~\textrm{kA/m} $~(149 mT). 
We consider an applied static field $H_x$ along the $x$ axis, resulting in a ``scissor'' state for the equilibrium magnetization which is given by $\mathbf{m}_\ell^\mathrm{eq}=H_x/H_j \hat{\mathbf{x}} + (-1)^{\ell+1} \sqrt{1- (H_x/H_j)^2}\hat{\mathbf{y}}$. 
%
The SAF has two eigenmodes~\cite{stamps_spin_1994}: the acoustic (ac) mode in which the dynamic, in-plane magnetization oscillate in phase, and the optical (op) mode in which these oscillations are in antiphase, as shown in the top inset of Fig.~\ref{GeneralDescription}(a). These eigenmodes can be computed from the dynamical matrix approach~\cite{grimsditch_magnetic_2004} and expressed~\cite{devolder_measuring_2022} as the complex vectors $\bm{\psi}_{\ell}^{\mathrm{ac, op}}=(\tilde m_{x\ell}, \tilde m_{y\ell}, \tilde m_{z\ell})$. We normalize the eigenvectors such that their creation corresponds to a decrement of one magnetization unit in each layer~\cite{korber_symmetry_2021}.


Figure~\ref{GeneralDescription}(a) presents the acoustic and optical mode frequencies as a function of the static applied field, $H_x$, with a schematic of the equilibrium state shown in the top inset.
The situation of primary interest, i.e., 3MS, involves the matching condition at which the frequency of the optical mode, $\omega_\mathrm{op} = \gamma_0  \sqrt{(H_j^2-H_x^2)M_s/H_j}$, is exactly twice that of the acoustic mode, $\omega_\mathrm{ac} = \gamma_0 H_x \sqrt{{(M_s+H_j)}/{H_j}}$, thereby ensuring energy conservation within the three-wave process. This occurs at the field $H_x=H_\mathrm{3MS} \equiv H_j \sqrt{(M_s)/(4 H_j+5 M_s)} \approx 0.44 H_j$. With our material parameters, $\omega_\mathrm{op} = 2 \omega_\mathrm{ac}$ is satisfied at $H_\mathrm{3MS}$= 51263 A/m (64.4 mT), corresponding to $\omega_\mathrm{op} / (2 \pi)=13.2$ GHz and  $\omega_\mathrm{ac} / (2 \pi)=6.6$ GHz. In the linear regime, the linewidths (FWHM) of these modes are $\Delta \omega_\mathrm{op} = {\alpha \gamma_0}  \left(M_s + (H_j^2-H_x^2)/(H_j^2) \right)=  {2 \pi} \times 582$ MHz and $\Delta \omega_\mathrm{ac} = {\alpha \gamma_0}  \left(M_s + (H_j^2+H_x^2)/(H_j^2)\right)=  {2 \pi} \times 600$ MHz. At this static field, $H_x=H_\mathrm{3MS}$, we also find that the threshold pumping field, $h_{\mathrm{rf},c}$, associated with a radiofrequency (rf) field applied along $x$, ${h_\mathrm{rf}} \cos ({\omega_\mathrm{rf}}\,t)$, is lowest when the pump frequency $\omega_\mathrm{rf}$ coincides with $\omega_\mathrm{op}$, as shown in Fig.~\ref{GeneralDescription}(b).

We now turn our attention to the dynamics of the acoustic and optical mode populations under different pumping levels. The eigenmode population is obtained by projecting of the dynamical magnetization $\mathbf{m}_\ell(t)$ onto the normalized eigenvector. Following ~\cite{daquino_novel_2009, perna_computational_2022, korber_symmetry_2021, massouras_mode-resolved_2024}, the complex mode amplitudes $\tilde b_\kappa(t)$ can be computed as $\tilde b_\kappa(t) = \sum_{\ell=1, 2} {(\bm{\psi}_{\ell,\kappa}^* \times \mathbf{m}_{\ell}^\mathrm{eq}) \cdot \big(\mathbf{m}_\ell(t)- \mathbf{m}_{\ell}^\mathrm{eq}\big)}$, where the index $\kappa$ refers to either the acoustic or optical mode. $\tilde b_\kappa(t)$ represents the classical counterpart to the expectation values of the magnon operators and are proportional to the transverse-component of the dynamic magnetization. We can define the time-dependent mode populations as $n_\kappa(t) = \tilde b_\kappa^* \tilde b_\kappa$.

The populations $n_\mathrm{op}(t)$ and $n_\mathrm{ac}(t)$ are expressed in units of normalized magnetization  {(i.e. the true number of magnons is $\frac 14 \frac{M_s} {\mu_B} \left(n_\textrm{op} + n_\textrm{ac} \right) V$ where $\mu_B$ is the Bohr magneton, $M_s/ \mu_B$ is the number of spins per unit volume and $V$ is the total volume of the SAF, see supplementary)}.

Below threshold, $h_\mathrm{rf} < h_{\mathrm{rf},c} = 389$~A/m, the optical mode population $n_\mathrm{op}(t)$ is pumped to a steady-state level, $n_\mathrm{op}^\infty= n_\mathrm{op}(t\rightarrow \infty)$, typically within a rise time of $2/\Delta \omega_\mathrm{op} \approx 1~\mathrm{ns}$, while the acoustic mode population $n_\mathrm{ac}(t)$ is exponentially damped to zero.
The threshold $h_{\mathrm{rf},c}$ is defined as the highest stimulus leading to $n_\mathrm{ac}^\infty = 0$. The numerical value found for $h_{\mathrm{rf},c}$ is consistent with the experimental observations~\cite{mouhoub_spin_2026}.

\begin{figure} 
	\centering\includegraphics[width=8.5cm]{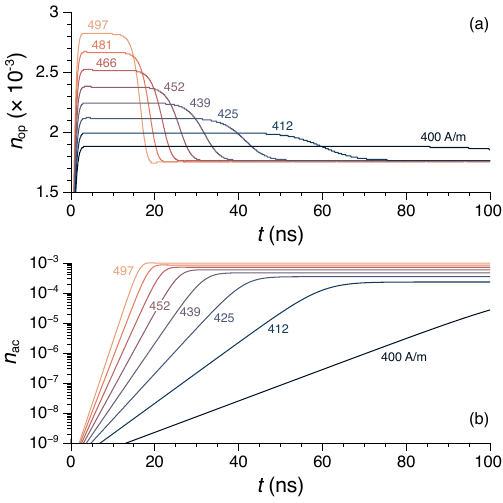}
	\caption{Population dynamics of the (a) optical and (b) acoustic mode for different rf field amplitudes, $h_\mathrm{rf}$, under the matching condition $\omega_\mathrm{rf}=\omega_\mathrm{op}$.}
	\label{TimeResolvedPopsAllRF}
\end{figure}

The population dynamics above threshold, $h_\mathrm{rf} > h_{\mathrm{rf},c}$, exhibits nontrivial features, as shown in Fig.~\ref{TimeResolvedPopsAllRF}. 
As before, the optical mode population attains a plateau $n_\mathrm{op}^\mathrm{max}$ within a similar rise time, but after an interval of a few tens of ns, the population decays towards a second plateau $n_\mathrm{op}^\infty$ which corresponds to the long-time behavior, as seen in Fig.~\ref{TimeResolvedPopsAllRF}(a). The transient plateau scales with the pumping field amplitude, 
\begin{equation}
\sqrt{n_\mathrm{op}^\mathrm{max}} \approx \chi_\mathrm{op} (h_\mathrm{rf}/{M_s}),
\label{susceptibilityOP}
\end{equation}
where $\chi_\mathrm{op} \approx 150$ represents the optical mode susceptibility to the pumping field. 
When $n_\mathrm{op}^\mathrm{max}$ is reached, $n_\mathrm{ac}$ grows exponentially as a result of 3MS, $n_\mathrm{ac}(t) = n_{\mathrm{ac},0}\exp\left(2\Gamma_\mathrm{ac} t\right)$, before also attaining a plateau, as shown in Fig.~\ref{TimeResolvedPopsAllRF}(b). $\Gamma_\mathrm{ac}^{-1}$ is the growth rate of the precession amplitude of the acoustic mode, which increases with $h_\mathrm{rf}$. Around the threshold region, this growth rate is proportional to the supercriticality parameter $\zeta \equiv h_\mathrm{rf}/h_{\mathrm{rf},c}$,
\begin{equation} 
\Gamma_\mathrm{ac} =  \Gamma_0 \, (\zeta -1). 
\label{SupercriticalityLinearFit}
\end{equation}
Fitting the data with this expression yields the decay rate extrapolated for zero stimulus, $\Gamma_0 = 1747~\mathrm{MHz}$ (see supplementary material). This is close to the linear relaxation rate expected from Gilbert damping, $\Delta \omega_\mathrm{ac}=1886~\mathrm{ MHz}$, where the discrepancy suggests that nonlinear frequency shifts of the mode is non-negligible even in the near-threshold regime.

At the transition between the two plateaus in $n_\mathrm{op}$, $n_\mathrm{ac}$ reaches a level comparable to that of $n_\mathrm{op}$ (Fig.~\ref{TimeResolvedPopsAllRF}). At this point, the sustained 3MS continues to scatter a sizable fraction of the optical modes such that $n_\mathrm{op}$ decreases, reducing in return the rate of creation of $n_\mathrm{ac}$. $n_\mathrm{ac}(t)$ is monotonic (no overshoot) only in the vicinity of the pumping threshold, i.e., $h_\mathrm{rf} \leq 450$~A/m where $n_\mathrm{ac}^\infty  = n_\mathrm{ac}^\mathrm{max}$. Outside of this regime, $n_\mathrm{ac}$ overshoots and $n_\mathrm{op}$ undershoots until the two populations stabilize at a fixed point. 
The final number of acoustic modes is proportional to the supercriticality (above threshold), 
\begin{equation}
n_\mathrm{ac}^\infty  \propto \zeta - 1.
\end{equation}
Note that $n_\mathrm{ac}^\infty$ scales linearly with $h_\mathrm{rf}$, in stark contrast to the direct excitation of $n_\mathrm{op}$ from the pumping field, which scales like $h_\mathrm{rf}^2$ (Eq.~\ref{susceptibilityOP}). The linear scaling of $n_\mathrm{ac}$ is reminiscent of parametric amplification, which in our case relates to the generation of $n_\mathrm{ac}$ by $n_\mathrm{op}$ through 3MS. It follows then that $n_\mathrm{ac}^\infty \propto  \Gamma_\mathrm{ac}$ for $\Gamma_\mathrm{ac} \geq 0 $, i.e., the final population of acoustic modes is simply proportional to their rate of creation.

We also note that $n_\mathrm{op}^\infty$ attains a ceiling at the threshold, which can be seen in Fig.~\ref{TimeResolvedPopsAllRF}(a) from the fact that the population values converge toward a single value at long times irrespective of $h_\mathrm{rf}$, i.e., $n_\mathrm{op}^\infty(h_\mathrm{rf}) \approx n_\mathrm{op}^\infty(H_\mathrm{3MS}^\mathrm{rf})$ for $\zeta \geq 1$. Above threshold, any excess in the pumping field level goes toward populating the acoustic mode. This phenomenon represents a special case in which self- and cross-terms in the nonlinear frequency shift balance out at the matching condition (see supplementary material). As we will see next, high supercriticalities, frequency detunings in the pumped field, or both, leads to more complex behavior.

\begin{figure}
	\centering\includegraphics[width=8.5cm]{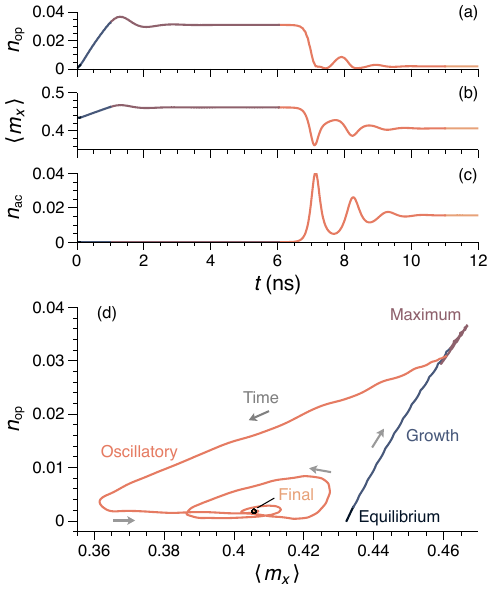}
	\caption{Population dynamics and correlation with the mean state at $H_x=  H_\mathrm{3MS}$ and a detuned stimulus frequency $\omega_\mathrm{rf}=12.68~\mathrm{GHz} < \omega_\mathrm{op}$ for an rf field of 2 kA/m. The initial state is the energy minimum. (a): Population $n_\mathrm{op}(t)$ of optical modes. (b) Period-averaged mean state $\langle m_x(t) \rangle$. (c) Acoustic population $n_\mathrm{ac}(t)$. (d) Correlation between the optical population and the period-averaged mean state. }  		\label{Pulling2000Am1268MHz}
\end{figure}

We now discuss how the pumped acoustic and optical mode populations affect the background magnetization state. To this end, we examine the period-averaged ``mean state'' which quantifies the dynamic equilibrium of the background magnetizations in the two layers. With $T = 2\pi/\omega_\mathrm{rf}$ representing the period of the stimulus, we define the mean state as the period-averaged magnetization component along the static field direction,
\begin{equation}
\langle m_x (t)\rangle = \frac{1}{2T} \int_{t}^{t+T} \left( m_{x,1}(t') + m_{x,2}(t') \right)dt' .
\label{MeanState}
\end{equation}  
Figure~\ref{Pulling2000Am1268MHz} shows the relationship between $\langle m_x \rangle$ and the mode populations for a detuned excitation, $\omega_\mathrm{rf}=12.68~\mathrm{GHz} < \omega_\mathrm{op}$, at a large driving field amplitude of $h_\mathrm{rf} = 2$~kA/m.
Over the first half of the interval shown, $n_\mathrm{op}$ reaches the initial plateau after a few ns [Fig.~\ref{Pulling2000Am1268MHz}(a)], mirrored by a similar change in the mean state $\langle m_x \rangle$ [Fig.~\ref{Pulling2000Am1268MHz}(b)], while $n_\mathrm{ac}$ remains essentially absent on the population scale considered [Fig.~\ref{Pulling2000Am1268MHz}(c)]. Within this initial phase, $n_\mathrm{op}$ is linearly correlated with $\langle m_x \rangle$, which can be discerned in the $(\langle m_x \rangle, n_\mathrm{op})$ phase plot in Fig.~\ref{Pulling2000Am1268MHz}(d) where the growth from equilibrium to the maximum value attained in $n_\mathrm{op}$ is traced out by a straight line. Over the second half of the interval $n_\mathrm{ac}$ attains a level comparable to that of $n_\mathrm{op}$, which is accompanied by large oscillations in $n_\mathrm{op}$, $\langle m_x \rangle$, and $n_\mathrm{ac}$ [Figs.~\ref{Pulling2000Am1268MHz}(a)-\ref{Pulling2000Am1268MHz}(c)]. More importantly, the mean state is no longer entirely locked to $n_\mathrm{op}$, but is also largely affected by $n_\mathrm{ac}$, as the oscillatory, spiralling trajectory toward the steady state shows in Fig.~\ref{Pulling2000Am1268MHz}(d).

Besides a short delay of $\approx$ 30 ps, we find that the mean state is well described by the empirical relation (see supplementary material),
\begin{equation}
\langle m_x \rangle = m_x^{\mathrm{eq}} + \left(\frac{\partial \langle m_x \rangle}{\partial {n_\mathrm{op}}}\right) n_\mathrm{ac} + \left(\frac{\partial \langle m_x \rangle}{\partial {n_\mathrm{ac}}}\right) n_\mathrm{op},
\end{equation}
which expresses the fact that the dynamical equilibrium $\langle m_x \rangle$ is determined by mutual pulling between the acoustic and optical mode populations. We find $\left(\partial\langle m_x \rangle/\partial n_\mathrm{op}\right)_{n_\mathrm{ac}=0} \approx 0.93$, the positive quantity expressing the fact that the optical modes compete with the antiferromagnetic interlayer interaction and favoring a more ferromagnetic alignment, while $\left(\partial\langle m_x \rangle/\partial n_\mathrm{ac}\right)_{n_\mathrm{op}=0} \approx -1.82$, the negative quantity expressing the fact that the acoustic mode favors rotating the total magnetic moment away from the applied field and favoring a more antiferromagnetic alignment. The opposite signs of the two pulling coefficients renders the nonlinear dynamics of the SAF fundamentally distinct from that in uniformly-magnetized films, where within nonlinear processes (notably Suhl instabilities~\cite{suhl_theory_1957}) the magnons at play \emph{all} pull the period-averaged mean state toward the same direction.

The period-averaged mean state $\langle m_x \rangle$ also influences the acoustic and optical mode frequencies away from equilibrium. To first order in $ \delta m_x = \langle m_x \rangle-m_x^\mathrm{eq}  \ll 1$ at $H_x=H_\mathrm{3MS}$, the mode frequencies are given by
\begin{subequations}
\begin{align} 
\omega_\mathrm{acNL} &\approx\omega_\mathrm{ac}+ \left(\frac{\omega_j}{2} \right) \delta m_x \label{NLomegaAc}, \\
\omega_\mathrm{opNL} &\approx\omega_\mathrm{op} - \left(\frac{3\omega_j}{4} \right) \delta m_x, \label{NLomegaOp}
\end{align}
\end{subequations}
where $\omega_j \equiv \gamma_0 M_s \sqrt{H_j/(H_j + M_s)}$ and the coefficient in front of $\delta m_x$ is termed the nonlinear frequency shift (NLFS). The opposite signs of the NLFS indicates that pulling of $\langle m_x \rangle$ increases the frequency of one mode while decreases the other. Moreover, the magnitude of these frequency shifts are also consequential compared to the Gilbert linewidth $\approx \alpha \gamma_0 M_s$ of the two modes. Since $\delta m_x$ can reach $\pm 0.05$ and $\sqrt{(H_j)/(H_j+M_s)} \approx 0.28$,  the NLFS coefficients can be of similar magnitude as the Gilbert linewidth ($\alpha =0.011$). Depending on the drive frequency $\omega_\mathrm{rf}$ and the applied static field $H_x$ (and consequently $\omega_\mathrm{op}$ and $\omega_\mathrm{ac}$), pulling of $\delta m_x$ can drive the modes in and out of resonance and modify the frequency detuning as the system evolves. As a result, a wide range of dynamical states can be accessed by varying the input parameters.

We can express the NLFS in terms of the mode populations. Using the pulling coefficients and eliminating the $\langle m_x \rangle$, we can see that the self-NLFS are both negative while the cross-NLFS are both positive, i.e.,
\begin{equation}
\frac{\partial \omega_\mathrm{ac}}{\partial n_\mathrm{ac}} <0, \, %
\frac{\partial \omega_\mathrm{op}}{\partial n_\mathrm{op}} <0, \, %
\frac{\partial \omega_\mathrm{ac}}{\partial n_\mathrm{op}} >0, \, %
\frac{\partial \omega_\mathrm{op}}{\partial n_\mathrm{ac}} >0.
\label{NLFS}
\end{equation}
All these quantities are of the same order of magnitude. Equation~(\ref{NLFS}) implies the following when resonantly pumping at the matching condition, $\omega_\mathrm{rf}=\omega_\mathrm{op}= 2\omega_\mathrm{ac}$, the NLFS pushes the system out of the linear resonance condition, resulting in $\omega_\mathrm{rf} \neq \omega_\mathrm{opNL}\neq  2\omega_\mathrm{acNL}$, which then limits the efficiency of the direct pumping of the optical mode by the rf field. However, frequency detuning of the input drive can compensate for the NLFS-induced detuning, such that the populations reached are higher compared to those obtained at the matching condition provided the threshold is overcome. Indeed, positive detuning $\omega > \omega_\mathrm{op}$ leads to higher thresholds, but once surpassed, higher population levels can also be attained (see supplementary material).

Since the transfer between the acoustic and optical mode populations occurs at a finite rate (related to $\Gamma_\mathrm{ac}$ or a similar quantity for the confluence), the cross-NLFS \emph{counteracts} the effect of the self-NLFS, but with a time delay. This results in the oscillatory behavior observed in Fig.~\ref{Pulling2000Am1268MHz}(d). Inspired from the population dynamics within two-magnon models of spin wave instability \cite{rezende_self-oscillations_1992, rezende_spin-wave_1993, de_aguiar_theory_2007}, we can exacerbate this competition between the NLFS coefficients to create a scenario in which the two populations never stabilize and the fixed point of the final populations transforms into a limit cycle through a subcritical Hopf bifurcation.

\begin{figure} 
	\centering\includegraphics[width=8.5 cm]{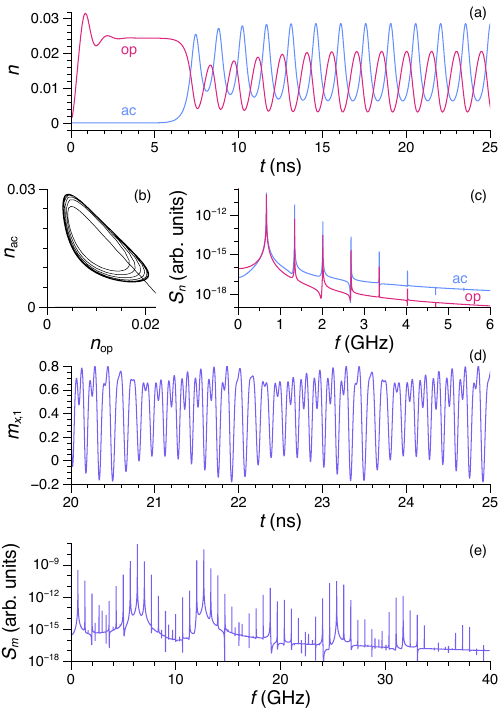}
	\caption{Self-induced Floquet modes. The field $H_x = 55~\mathrm{kA/m} > H_\mathrm{3MS}$ is larger than the matching condition. A non-resonant stimulus is applied with $\omega_\mathrm{rf} = 12.68\mathrm{~GHz} < \omega_\mathrm{op}  < 2\omega_\mathrm{ac}$ is applied, with a peak amplitude $h_\mathrm{rf}=2.2$~kA/m (2.8 mT). (a) Time dependence of the acoustic and optical mode populations. (b) Population phase plot showing limit cycle. (c) Power spectral density of the mode population, $S_n$, exhibiting a characteristic frequency at 672 MHz. (d) Time dependence of $x$-component of the magnetization of the first layer $m_{x1}(t)$ within the limit cycle. (e) Power spectral density of $m_{x1}$, $S_m$, showing the frequency combs around $\omega_\mathrm{rf}/2$ and its harmonics.}
	\label{Sweet spot}
\end{figure}

Figure~\ref{Sweet spot} presents such a scenario, which is obtained with $f_\mathrm{rf} =  12.68~\mathrm{GHz} < f_\mathrm{op}= 12.98~\mathrm{GHz}  <  2f_\mathrm{ac} = 14.17~\mathrm{GHz}$ and $h_\mathrm{rf}=2.2$~kA/m (2.8 mT).  
Intially, the frequency detuning between the pump and the optical mode is small such that the NLFS acts to drive the optical mode out of resonance. After a transient phase that resembles the previous cases, the $n_\mathrm{op}$ periodically decays (as a result of 3MS and dynamic detuning) and grows (through the rf field pumping), with $n_\mathrm{ac}$ exhibiting a similar pattern with a phase lag. The system ultimately settles into a limit cycle, as shown in Fig.~\ref{Sweet spot}(b), where the subsequent dynamics of the two mode populations is periodic. The power spectra of $n_\mathrm{op}$ and $n_\mathrm{ac}$, $S_n(\omega)$, are shown in Fig.~\ref{Sweet spot}(c), from which we can deduce the characteristic period of  1.49 ns corresponding to a frequency of ${\omega_\mathrm{cycle}}/{2 \pi}= 672~\mathrm{MHz}$. \textcolor{black}{With our material parameters, periodic oscillations of the populations could be induced for damping values up to  $\alpha > 0.018$ but not above, even if increasing the stimulus amplitude (see supplementary).}

The cycle of growth and decline of the two populations is reminiscent of predator-prey dynamics described by the Lotka-Volterra equations. Here, the rf field pump represents an external source of prey (the optical modes), which naturally decay at a rate $\Delta \omega_\mathrm{op}$. Predators (the acoustic modes) thrive by consuming these prey through 3MS when prey is plentiful, while naturally decaying at the rate $\Delta \omega_\mathrm{ac}$. The Lotka-Volterra model, which does not account for the additional complexity arising from the NLFS, predicts $\omega_\mathrm{cycle} = \sqrt{\Delta \omega_\mathrm{ac} \Delta \omega_\mathrm{op}}/{2 \pi}= 591~\mathrm{MHz}$, in relatively good agreement with our numerical result that fully incorporates the NLFS.

The key result here is that the oscillatory dynamics of the optical and acoustic mode populations induce in turn a time-periodic modulation of the dynamic equilibrium, $\langle m_x \rangle$, which is initiated by pumping of the eigenmodes of the system. As mentioned in the introduction, this consititutes another example of self-induced Floquet states, whereby pumping of an eigenmode results in the time-periodic modulation of the background state, through nonlinear interactions, which subsequently induces Floquet states. An example of the time-domain magnetization at steady state is shown in Fig.~\ref{Sweet spot}(d), which exhibits several oscillations reflecting the three characteristic frequencies of the driven system: the limit cycle $\omega_\mathrm{cycle}$ and the two forced mode frequencies $\omega_\mathrm{rf}/2$ and $\omega_\mathrm{rf}$. The population dynamics are intertwined with the mode-mode coupling such that the power spectral density of the magnetization, $S_m(\omega)$, exhibits a frequency comb with peaks at every harmonic of $\omega_\mathrm{rf}/2$, each of which with sidebands spaced at multiples of $\omega_\mathrm{cycle}$. All harmonics of $\omega_\mathrm{cycle}$ are also present, but since $\omega_\mathrm{cycle}$ is unrelated to the forcing frequency $\omega_\mathrm{rf}$, the peaks at $n \omega_\mathrm{cycle},~n\in \mathbb N$ adds to the magnonic frequency comb at $v \omega_\mathrm{rf}/2 + w \omega_\mathrm{cycle}$, with $v\in \mathbb N^*,~w \in \mathbb Z^*$. This doubles the comb density at the edge of the Floquet Brillouin zone, i.e., near $\frac 14 \omega_\mathrm{rf}=2 \pi \times 3.17$ GHz in Fig.~\ref{Sweet spot}(e).

Since $\omega_\mathrm{cycle}$, $\Delta \omega_\mathrm{ac}$ and $\Delta \omega_\mathrm{op}$ are of the same order of magnitude ($2 \pi \times 600~\mathrm{MHz}$), the populated acoustic and optical modes remain coherent over several periods of the induced modulation characterized by $\omega_\mathrm{cycle}$. As a consequence, we can consider the comb modes in Fig.~\ref{Sweet spot}(e) as Floquet modes arising from the time-periodic modulation of the system Hamiltonian. This behavior also contrasts with the self-modulated spin-wave instability observed in ferromagnetic systems, where the modulation results from four-magnon scattering and is typically much too slow (MHz range, \cite{rezende_model_1986, rezende_self-oscillations_1992}) to drive Floquet magnons. \textcolor{black}{Compared to the self-induced Floquet states involving vortex gyration~\cite{heins_self-induced_2026}, the nature of the period-averaged ground state here on which the Floquet modes reside cannot be anticipated, neither from the linear eigenstates of the system, nor from the linear response to the drive.}

Our model here considers only spatially-uniform acoustic and optical modes, i.e., at zero wave vector, which differs from the 3MS processes observed experimentally involving finite wavevectors in extended systems~\cite{mouhoub_spin_2026}. The key ingredient for generating the magnonic frequency comb in our work is the 3MS involving two modes, with no substantial coupling to other modes and a large feedback of the populations on their growth rates. A practical realization of the self-induced mechanism within a SAF could, for instance, involve confined geometries like sufficiently small SAF dots as shown in the supplementary. This might be achieved in magnetic tunnel junctions with a SAF free layer.

We acknowledge financial support from the EU Research and Innovation Programme Horizon Europe under grant agreement n°101070290 (NIMFEIA).

\pagebreak
\renewcommand{\theequation}{S\arabic{equation}}
\renewcommand{\thefigure}{S\arabic{figure}}
\renewcommand{\bibnumfmt}[1]{[S#1]}
\renewcommand{\citenumfont}[1]{S#1}
\setcounter{figure}{0}

%
\begin{widetext}
\begin{center}
\textbf{\large Supplementary: Self-induced Floquet states via three-wave processes in synthetic antiferromagnets}
\end{center}
\end{widetext}

%
%

Abstract: \textit{This is supplementary of the paper "Self-induced Floquet states via three-wave processes in synthetic antiferromagnets". It provides details about the analytical methods used for the analysis of the time-domain magnetizations. It then reports examples of the time-evolution of the total energy of the system of the total number of quasiparticles hosted by the system. The dependence of the maximum transient populations, and of the final populations of the acoustic and optical modes are reported, first versus the applied rf field amplitude, then versus also its detuning away from the optical mode. The analysis of the torque supplied by the optical mode on the acoustic mode is used to discuss the difference between the generation of acoustic modes by three-magnon scattering or by direct parametric pumping from the rf applied field. The self non-linear frequency shift and that across modes are used to understand how to maximize the number of magnons hosted by the system. \textcolor{black}{The influence of the initial magnetic state is discussed, as well as the choice of the detuning conditions that yield to the self-induced Floquet dynamics. Finally, this dynamics is implemented in a SAF nanodot described by full micromagnetics.}}

This supplementary is organized as follows. \\
Section~\ref{analytics} gives further details about the nature of the synthetic antiferromagnet (SAF) eigenmodes, their normalization, the methods used to follow their populations and the population levels when at thermal equilibrium. \\
Section~\ref{additionalMetrics} discusses the evolutions of the number of magnons resulting in a situation where the resonant inductive pumping of the optical mode leads to three-magnon scattering at the perfectly matching condition. \\
Section~\ref{StoredEnergy} analyses the energy stored in the eigenmode system in similar conditions as section ~\ref{additionalMetrics}. \\
Section~\ref{populationExtrema} discusses the dependence of the transiently overshooting populations and of the final populations when the applied frequency is resonant with the optical mode and when the field complies with the perfect matching condition.\\
Section~\ref{MeanState} demonstrates that the period-averaged mean state at a given time can be deduced from the sole knowledge of the populations of the two modes at this same given time. \\
Section~\ref{AmplificationRate} gives further details about the dependence of the amplification rate of the acoustic population versus the amplitude of the stimulus. \\
Section~\ref{3MSversusPP} discusses the physical origin of three-magnon splitting (3MS) in SAFs by comparing it with situations in which the acoustic mode would be parametrically amplified directly by the applied rf field. \\
Section \ref{NLomegas} reports the non-linear eigenmode frequencies for any applied field leading to a scissors state. \\
Section~\ref{PopulationMaximization} finally discusses how to detune the applied frequency to maximize the obtained magnon populations.\\
Section~\ref{PhaseDiagrams} studies the size of the parameter space leading to either self-induced Floquet states or steady populations.\\
\textcolor{black}{Section~\ref{InfluenceInitialState} discusses how robust is the Floquet dynamics versus choices of the initial magnetizations. \\
Section~\ref{Detuning} discusses how to detune the applied field and the applied frequency away from the perfectly matching condition in order to enter the dynamical regime of cyclic growth and decay of the two mode populations.} \\
Section~\ref{micromagnetics} finally shows an example of self-induced Floquet state obtained in a realistic sample described in the framework of micromagnetism.

\section{Analytics of the dynamical system: eigenpairs and populations} \label{analytics}

\subsection{Normalized eigenmodes}
A SAF in the two-macrospin approximation has two eigenmodes: the acoustic one (ac) in which the scissors rigidly rocks and the optical one (op) in which the scissors breathes [sketchs in the figure 1 of the main document]. These eigenmodes can be deduced from the dynamical matrix approach \cite{grimsditch_magnetic_2004} and expressed as 6-component complex vectors $\{\tilde m_{x1}, \tilde m_{y1}, \tilde m_{z1}, \tilde m_{x2}, \tilde m_{y2}, \tilde m_{z2}\}$, as can be found in ref.~\cite{devolder_measuring_2022}. In the layer $\ell$, the acoustic mode has the following dynamical magnetization:
$$\vec \psi_{\ell}^{\textrm{ac}}= \mathcal B_\textrm{ac} {\left(\bar \ell  i \sqrt{\frac{M_s+H_j}{H_j}}\frac{\sqrt{H_j^2-H_x^2}}{H_x},+i   \sqrt{\frac{M_s+H_j}{H_j}}, 1 \right) }$$ 
where $\bar \ell = (-1)^\ell$ and $i$ is the imaginary unit. The $\vec \psi_{\ell}^{\textrm{ac}}$'s are perpendicular to the $\vec m_\ell^\textrm{eq}$'s. Similarly, the dynamical magnetization of the optical mode are:
$$ \vec \psi_{\ell}^\textrm{op}=  \mathcal B^{\textrm{op}} {\left(+i \sqrt{\frac{M_s}{H_j}},  \bar \ell i \sqrt{\frac{M_s}{H_j}} \frac{H_x    }{\sqrt{H_j^2-H_x^2}}, \bar \ell  \right) }$$ 
Being eigenvectors, the $\vec \psi$'s are defined up to a multiplicative number. Following \cite{korber_symmetry_2021}, we normalise the $\vec \psi$'s so that their creation/annihilation correspond to an decrement/increment of one unit of longitudinal magnetization in each layer layer: we define the complex amplitudes associated with the eigenvectors $\vec \psi$'s as:
\begin{equation} 
\tilde b_{\psi}^\textrm{mode} =  {\sum}_{\ell} \big( \vec m_\ell^\textrm{eq}\times \vec \psi_\ell^\textrm{mode} \big).(\vec \psi_\ell^\textrm{mode})^*
\end{equation} 
The prefactors $\mathcal B^{\textrm{ac}}$ and $\mathcal B^{\textrm{op}}$ can be determined by writing that the complex mode amplitudes $\tilde b_{\psi}^\textrm{mode}$ of each eigenvector correspond to one unit of magnetization per layer in the direction $\vec m^\textrm{eq}$, which is done by writing $\tilde b_{\psi}^* \tilde b_{\psi} = 1$. With our material parameters and field conditions, the normalization prefactors will be $\mathcal B^{\textrm{ac}} \approx \frac16$ and $\mathcal B^{\textrm{op}}\approx \frac 14$.

\subsection{Following the number of magnons for each eigenmode} 
We aim to study the dynamics of the populations of each eigenmode;  therefore the term population deserves to be clearly defined. A  straightforward measurement of the population could be given by a demodulation of the sums or differences of the magnetizations of the two layers, i.e.: 
$$\Big \Vert \int_{t}^{t+ 2\pi/\omega_\textrm{mode}} \big(m_{z1}(t') \pm m_{z2}(t') \big) e^{i \omega_\textrm{mode} t'} dt' \Big \Vert $$
where the $\pm$ is $+$ for the acoustic mode and $-$ for the optical mode. Such a definition would suffer a major deficiency: each populations would be defined with its own arbitrary scale and would not be informative on the number of magnons in the mode.  

Instead, we follow the method of refs.~\cite{daquino_novel_2009, korber_symmetry_2021, massouras_mode-resolved_2024}, based on the projection of the dynamic part of the magnetizations onto the normalized eigenmodes.
We therefore compute mode-resolved mode complex amplitudes from the time-dependent magnetizations: 
\begin{equation}
\tilde b_\textrm{mode}(t) = \sum_{\ell=1, 2} {(\vec \psi_{\ell,\textrm{mode}}^* \times \vec m_{\ell}^\textrm{eq}).\big(\vec  m_\ell(t)-\vec m_{\ell}^\textrm{eq}\big)} ~\in \mathbb C
\label{DemodulationProcedure} 
\end{equation}
The $\tilde b_\textrm{mode}(t)$ are the classical analogs of the expectation values of the magnon operators and are proportional to the transverse-precessing magnetizations. We then calculate the time-resolved populations:
\begin{equation}
n_\textrm{mode}(t)    = \tilde b_\textrm{mode}^* \tilde b_\textrm{mode}  ~\in \mathbb R^+
\label{DefinitionPopulations}
\end{equation}
which are the classical analogs of the magnon occupancy numbers. These populations $n_\textrm{op}(t)$ and $n_\textrm{ac}(t)$ are expressed in units of normalized (longitudinal) magnetization along the equilibrium direction (i.e. in units of half a magnon per spin, or equivalently in spin per spin). The transverse components of the mode-resolved dynamical magnetization (the mode-resolved "precession amplitudes") are proportional to the quantity $\sqrt n_\textrm{mode} = || \tilde b_\textrm{mode} ||$.

\subsection{Thermal seed to initiate the three-magnon scattering process}
The growth of the population of the acoustic mode by 3MS ressembles a parametric amplification process induced by the dynamic interlayer exchange field supplied by the optical mode. This amplification process requires a finite fluctuation --a thermal seed-- to initiate. Setting initial conditions ($t=0$) that depart from the minimum energy state ensures a more realistic description of the population dynamics. Using Table 1 of ref. \onlinecite{devolder_measuring_2022}, we will mimic an initiallly thermal population of the acoustic mode by performing a small in-plane rotation of the magnetizations of the two layers away from equilibrium:
\begin{equation} \delta \phi_{1, 2} =  \sqrt{\frac{k_B T}{2 \mu_0 H_j  M_s V (1-{{H_x}^2}/{{H_j}^2})}} \label{ThermalFluctuationAmplitude} \end{equation} 
where $V$ is the volume in which we describe the dynamics. For the forthcoming numerical evaluations, we shall take a volume $V = 2\times 17~ \textrm{nm} \times 30 ~ \mu \textrm{m}^2 $ (i.e. $ n_\textrm{spins} \approx1.5\times 10^{11}$ spins). This yields $\delta \phi_{1, 2}\approx0.007^\circ$ at room temperature, or equivalently $n_\textrm{ac}(t=0,~300~\textrm{K}) \approx 10^{-9} $. 

%
\begin{figure} 
{\includegraphics[width=9cm]{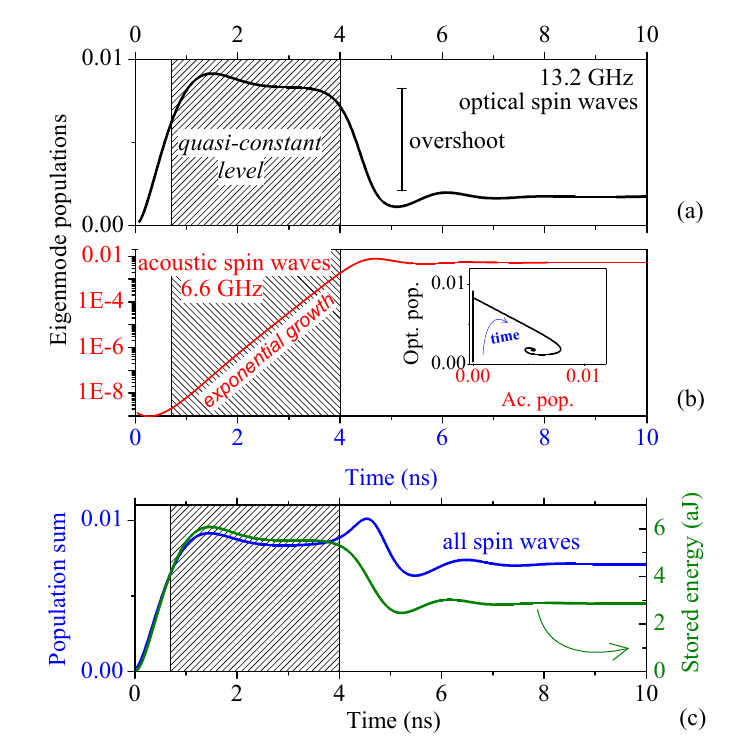}}
{\caption{Population dynamics for an rf field $H_x^\textrm{rf}=1~\textrm{kA/m}$ and perfectly matching conditions when starting from the thermal seed and resonantly pumping the optical mode. (a) Time-resolved population population of the optical mode. (b) Idem for the acoustic mode. Inset: population-population plot. (c) Sum of the two populations (blue) and total energy of the system (green). }\label{TimeResolvedPopulations1kAperMeter}}
\end{figure}

\section{Total number of magnons and energy stored in the eigenmodes}\label{additionalMetrics}
It is interesting to look at additional metrics of the dynamics, as illustrated in Fig.~\ref{TimeResolvedPopulations1kAperMeter} in perfectly matching conditions when resonantly pumping the optical mode. 
A first interesting metrics is the total population of modes $n_\textrm{tot}(t)=n_\textrm{op}+ n_\textrm{ac}$  [blue curve in Fig.~\ref{TimeResolvedPopulations1kAperMeter} (c)]. In a 3-magnon splitting (3MS) process, the annihilation of one optical magnon comes with the creation of two acoustic magnons, so: 
$$dn_\textrm{ac}\big{\lvert}_\textrm{3MS} = -2 dn_\textrm{op}\big{\lvert}_\textrm{3MS}$$
We could therefore expect that the total population of magnons would increase like $dn_\textrm{tot}\big{\lvert}_\textrm{3MS} = \frac 12 dn_\textrm{ac}\big{\lvert}_\textrm{3MS}$ as a result of 3MS, up to potentially doubling if all optical magnons were to split and no acoustic magnon were to undergo confluence. 
This is not the case: only a transient increase of the total population is seen when 3MS gets frequent [at $t \in [4, 5]$ ns in the representative example of Fig.~\ref{TimeResolvedPopulations1kAperMeter}(c)]. This is due to a combination of several effects: the start of the reverse process (confluence), and more importantly the alteration of the susceptibility of the optical mode though its non-linear frequency shift, which makes it effectively frequency-detuned from the stimulus frequency and there decreases its rate of creation by the external rf field. This is confirmed by simulations (not shown) done with zero damping and no rf field, but starting from random populations levels which systematically gives results that comply with the fluxes of quasi-particules expected when 3MS is the sole mechanism at play (Eq.~\ref{MagnonNbNonConservation}). In this $\alpha=0$ and $H_x^\textrm{rf}=0$ very unrealistic case, $n_\textrm{ac} + 2 n_\textrm{op}$ is a conserved quantity during the time evolution.

\section{energy stored in the eigenmodes}\label{StoredEnergy}
In the absence of (non-linear) mode-mode coupling, the energy contained in the eigenmodes can be expressed as that of a sum of harmonic oscillators with an energy density (in Joule/m$^3$) being:
\begin{equation}  
 E_\textrm{modes}  = \frac 14 \frac{M_s} {\mu_B} \left(n_\textrm{op} \hbar \omega_\textrm{op} + n_\textrm{ac} \hbar \omega_\textrm{ac}\right) \label{MagnonNbNonConservation}  \end{equation} 
 where $\mu_B$ is the Bohr magneton, $M_s/ \mu_B$ is the number of spins per unit volume  {and the factor $\frac 14$ corrects for the number of layers and the fact that a magnon is a change of to units of spin}. Provided the energy of the equilibrium state is conventionally taken as the zero, this definition yields almost exactly the same result as the direct computation of the energy density of the system, as calculated directly from the time-dependent magnetizations, i.e.:
 \begin{multline}
 E_\textrm{tot}= \frac 12 \times \Big[\\ - \mu_0 M_s \left( (H_x+H_x^\textrm{rf} \textrm{cos}(\omega_\textrm{rf} t)) .(m_{x1}+m_{x2}) \right) \\
  \frac J t \vec m_1.\vec m_2 + \frac{\mu_0 M_s^2}{2} \left(  m_{z1}^2+m_{z2}^2 \right) \Big]
  \label{energyLLG}
\end{multline}

The very small difference between $E_\textrm{tot}$ and $E_\textrm{modes}$ arises from the Zeeman energy of the (small) rf field, from a slight difference between the equilibrium state and the time-averaged mean state at large pumping powers and from the non-linear change of the eigenmode frequencies. This quasi-equality of $E_\textrm{tot}$ and $E_\textrm{modes}$ also means that the higher order terms of the type $\tilde b_\textrm{ac}^* \tilde b_\textrm{op}$ do not substantially modify the energy of the system in the range parameters investigated here.

The energy stored in the eigenspectrum is also an interesting metrics of the dynamics. In [Fig.~\ref{TimeResolvedPopulations1kAperMeter}(c)] it appears that the transfer of population from the optical mode to the acoustic mode reduces the ability of the system to absorb the power supplied by the stimulus field. This can be understood from the result of section~\ref{populationExtrema}: the cross NLFS suppresses the resonant character of the optical mode with the pump when the acoustic modes reaches a sizable population level. 

\begin{figure} 
\includegraphics[width=9cm]{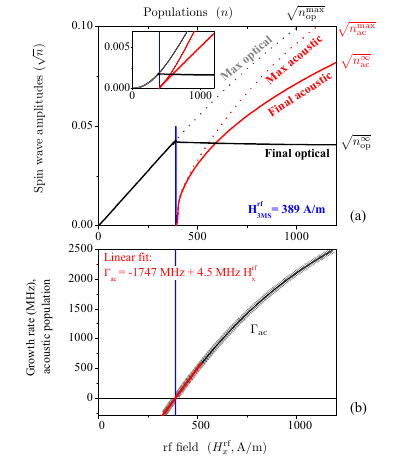} 
\caption{Population metrics in perfectly matching conditions: (a) Final $\sqrt{n^\infty}$ and maximum transient $\sqrt{n^\textrm{max}}$ amplitudes for the acoustic (red) and optical  (black) modes. Inset: same data but in population scale. (b) Rate of growth of the amplitude of the acoustic mode. The red dotted line is a linear fit through the near threshold region.} \label{Fig_Final_Max_Pops_and_Growth_Rate}
\end{figure}


\section{Population extrema in perfect matching condition and resonant pumping} \label{populationExtrema}

The Fig.~\ref{Fig_Final_Max_Pops_and_Growth_Rate} reports the final mode populations $n_\textrm{ac, op}^\infty$ as well their (transient) maximum levels $n_\textrm{ac, op}^\textrm{max}$ when the system is prepared in perfectly matching condition with $H_x=H_\textrm{3MS}$, and it is resonantly excited with $\omega_\textrm{rf}=\omega_\textrm{op}$ from a thermal seeded initial state (Eq.~\ref{ThermalFluctuationAmplitude}).

\subsubsection{Transient and final populations of the optical mode}\
From Fig.~\ref{Fig_Final_Max_Pops_and_Growth_Rate}(a), the optical populations have several noticeable properties. 
The first one is that there is overshoot of the optical population if and only if we are above the 3MS threshold:
\begin{equation}
n_\textrm{op}^\infty =n_\textrm{op}^\textrm{max} ~~~\Leftrightarrow ~~~~~ H_x^\textrm{rf} \leq H_{3MS}^\textrm{rf}= 389 \textrm{~A/m} 
\end{equation}
In addition, the maximum (transient) of the amplitude of the optical modes scales with the stimulus: 
\begin{equation}
\sqrt {n_\textrm{op}^\textrm{max}}  ~ \approx \chi_\textrm{op} \frac{H_x^\textrm{rf}} {M_s}, ~~~~~~~~\forall H_x^\textrm{rf} \label{susceptibilityOP}
\end{equation}
The linearity of the Eq.~\ref{susceptibilityOP} recalls that the optical mode is transiently excited thanks to a conventional susceptibility. Eq.~\ref{susceptibilityOP} expresses simply the conversion yield between the applied microwave photons --their number scales with ${\mu_0 (H_x^\textrm{rf})^2}/(\hbar \omega_\textrm{rf} )$-- and the transiently obtained optical modes (their number being $n_\textrm{op}^\textrm{max}$). In  Eq.~\ref{susceptibilityOP}, the proportionality holds \textit{strictly} when below the threshold (a linear fit has a Pearson coefficient of exactly 1 and yields $\chi_\textrm{op}$ = 149.6, in close agreement with the transverse susceptibility \footnote {Using a first order expansion in $\alpha$ of the susceptibility of SAFs and the normalization of the eigenvectors, after some algebra this number can be rewritten as $\chi_\textrm{op} \approx \sqrt{\frac{M_s} {H_j}} \frac {2 \omega_\textrm{op}}{\Delta \omega_\textrm{op}}$, which would give a value 153 instead of the measured 149.6 }). 

In contrast, above the threshold, the maximum attained optical amplitude is slightly less that extrapolated from the susceptibility [see Fig.~\ref{Fig_Final_Max_Pops_and_Growth_Rate}(a)]: we have $\sqrt {n_\textrm{op}^\textrm{max}}  \leq \chi_\textrm{op} {H_x^\textrm{rf}} / {M_s}$. Eq.~\ref{susceptibilityOP} becomes sub-linear. Note that $\sqrt {n_\textrm{op}^\textrm{max}}$ is already defined at short times, much before the energy of the optic mode flows away to the acoustic population by 3MS. The effective reduction of $\chi_\textrm{op}$ does thus not arise from 3MS but rather arises from the self non-linear frequency shift (NLFS) of the optical modes that make them detuned from the stimulus frequency. We will come back this point in the last section.

Finally, in a striking manner, the final optical population $\sqrt {n_\textrm{op}^\infty}$ hits a hard roof at the threshold (it "clips") and then stays essentially constant \footnote{More precisely, our calculations indicate that the final optical population is not constant but decreases slowly over the range of studied stimuli. We will see later that this somewhat counterintuitive (''more stimulus, less response'') clipping character of $n^{\infty}_\textrm{op}$ versus $H_x^\textrm{rf}$ depends on the cross non-linear frequency shift that the acoustic mode amplitude exerts on the frequency of the optical mode.}.
In perfectly matching condition and resonant pumping, we have: 
\begin{equation}
n_\textrm{op}^\infty(H_x^\textrm{rf}) \approx n_\textrm{op}^\infty(H_\textrm{3MS}^\textrm{rf})~~~\textrm{for~~} H_x^\textrm{rf} \geq H_\textrm{3MS}^\textrm{rf} \label{clippingOP}
\end{equation}

\subsubsection{Transient and final populations of the acoustic mode}
From Fig.~\ref{Fig_Final_Max_Pops_and_Growth_Rate}(a), the acoustic population have several noticeable properties. When below the threshold the rate of annihilation of the acoustic modes by Gilbert damping exceeds the rate of their creation by 3MS, such that in the end:
\begin{equation}
n_\textrm{ac}^\infty  = 0 ~~~\textrm{for~} H_x^\textrm{rf} \leq H_\textrm{3MS}^\textrm{rf} 
\end{equation}
In addition, Fig.~\ref{Fig_Final_Max_Pops_and_Growth_Rate}(a) shows that there is absence of overshoot of the acoustic population uptill a bit (1.3 dB) above the threshold:
\begin{equation}
n_\textrm{ac}^\infty  =n_\textrm{ac}^\textrm{max} ~\Leftrightarrow ~ H_x^\textrm{rf} \leq 450  \textrm{~A/m} 
\end{equation}
Lastly, the final number of acoustic modes is proportional to the supercriticality:
\begin{equation}
n_\textrm{ac}^\infty  \propto \frac{H_x^\textrm{rf}-H_{3MS}^\textrm{rf}}{H_{3MS}^\textrm{rf}} ,~~~~~\textrm{for~} H_x^\textrm{rf} \geq H_\textrm{3MS}^\textrm{rf} \label{linearityACpop}
\end{equation}
Noticeably, the proportionality is now between $n_\textrm{ac}^\infty$ and a field $(H_x^\textrm{rf}-H_{3MS}^\textrm{rf})$, in start contrast with the case of excitation by conventional susceptibility (Eq.~\ref{susceptibilityOP}) for which $n_\textrm{op}^\textrm{max} \propto (H_x^\textrm{rf})^2$. We will see that this is reminiscent of a parametric amplification situation.

\section{Period-averaged mean state}\label{MeanState} 
\begin{figure} 
{\includegraphics[width=8 cm]{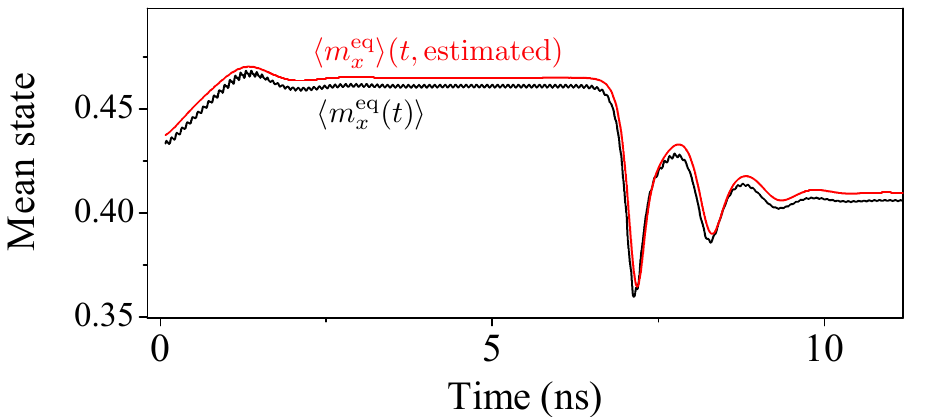}}
{\caption{Comparison of the period-averaged mean state (black) with its estimation (red) from the populations of the two eigenmodes according to Eq.~\ref{JoinedPulling} from the data of Fig.~3 of the main document. Since the curves are essentially superposed, we have offset the red curve vertically to improve readability.}  \label{recalculatedGroundState}}
\end{figure}

The period-averaged mean state, as defined in the main document, can be understood solely from the joint pulling actions of the two populations. The quantity:
\begin{equation}
{\langle m_x\rangle} (t, \textrm{estimated})= m_x ^\textrm{eq} +  \frac {d\, {\langle m_x }\rangle} {d\, {n_\textrm{op}}} n_\textrm{ac}   
+
\frac {d\, {\langle m_x }\rangle} {d\, {n_\textrm{ac}}}  n_\textrm{op}  
\label{JoinedPulling}
\end{equation}
is an estimation of the mean state deduced from the populations. Fig.~\ref{recalculatedGroundState} compares it with the true mean state. Apart from a small time delay ($\approx$ 30 ps) expressing that the pulling is the \textit{consequence} of the creation of magnons, the agreement is remarkable. The mean state is thus a quick-responding function of the populations.

\section{Rate of amplification of the acoustic mode} \label{AmplificationRate}
The main document defines $\Gamma_\textrm{ac}^{-1}$ as the rise time of the precession amplitude of the acoustic mode whenever the level of the optical population is rather constant. 
The stimulus dependence of $\Gamma_\textrm{ac}$ is plotted in Fig.~\ref{Fig_Final_Max_Pops_and_Growth_Rate}(b) for perfectly matching condition and resonant pumping. The growth/decay rate seems to depend linearly on the stimulus amplitude, at least in the region of the threshold. We therefore fit  $\Gamma_\textrm{ac}$ versus $H_x^\textrm{rf}$ with a supercriticality law that mimics that observed in many threshold phenomena:
\begin{equation} 
\Gamma_\textrm{ac} =  \Gamma_0 \times \frac{H_x^\textrm{rf} - H_{3MS}^\textrm{rf}} {H_{3MS}^\textrm{rf}} \label{SupercriticalityLinearFit}
\end{equation}
A fit with this expression is done in Fig.~\ref{Fig_Final_Max_Pops_and_Growth_Rate}(b). It yields a slope of $4.5~\textrm{MHz.m.A}^{-1}$ and a  decay rate extrapolated for zero stimulus that is $\Gamma_0 = 1747\pm 3~\textrm{MHz}$. In absence of NLFS, this number should be equal to the Gilbert linewidth of the acoustic mode $\Delta \omega_\textrm{ac}=1886~\textrm{ MHz}$.

\section{Three magnon splitting versus direct parametric pumping}\label{3MSversusPP}
Several features of 3MS in SAF --the presence of a threshold, the proportionality of the final acoustic population with the supercriticality, the shape of the threshold versus frequency curve \cite{srivastava_identification_2023}-- bear a strong similarity with situations of parametric amplification. One may legitimately wonder if the presently obtained dynamics is really the result of 3MS from optical magnons to pairs of acoustic magnons, or whether parametric pumping of the acoustic mode \textit{directly} from the rf field could be the key phenomenon. Indeed the acoustic mode is essentially the precession of the \textit{sum} of the magnetizations of the two layers and this precession is elliptical. After all, a macrospin magnet having $(\vec m_1 + \vec m_2)$ as magnetization submitted to an rf field at the frequency $2\omega_\textrm{ac}$ would undergo parametric amplification of its Kittel mode and no optical mode would need to be invoked. 
This hypothesis of direct parametric pumping can be easily excluded by additional calculations made in parametric pumping conditions $\omega_\textrm{rf}=2\omega_\textrm{ac}$ but far from the perfectly matching condition. Such a configuration does not excite the acoustic mode up to unreasonably large rf fields. 

The similarity of the presently reported dynamics of 3MS in SAF with parametric amplification situations can be qualitatively understood by looking at the microwave effective field that an optical mode would apply on a macrospin having $(\vec m_1 + \vec m_2)$ as magnetization. 
The \textit{sum} of the effective fields $\vec H_\textrm{eff}^\textrm{sum}= \vec H_\textrm{eff}(\vec \psi_\textrm{op})^{\ell=1}+\vec H_\textrm{eff}(\vec \psi_\textrm{op})^{\ell=2}$ that an optical mode $\vec \psi_\textrm{op}$ applies on the \textit{summed} magnetizations of the two layers can be deduced from the expression of $\vec \psi_\textrm{op}$ and the effective field steming from the gradient of the total energy (Eq.~\ref{energyLLG}). This sum of effective fields is:
\begin{equation} \vec H_\textrm{eff}^\textrm{sum} =  {\tilde b_{\textrm{op}}} \mathcal B_\textrm{op} (2 H_\textrm{3MS} - i \sqrt{H_jM_s}) \vec e_x \label{EffectiveFieldFromOpticalMode}
\end{equation}
This sum is along $\vec e_x$ and therefore has the right orientation and the right frequency to parametrically pump the Kittel mode of macrospin having $\vec m_1 + \vec m_2$ as magnetization. Using our material parameter 
and Eq.~\ref{susceptibilityOP}, we can see that 
$|| \vec H_\textrm{eff}^\textrm{sum}  || \approx 11.5 || \vec H_x^\textrm{rf} ||$, i.e. it is much larger than the applied rf field and therefore plays by far the major role for populating of the acoustic mode. Qualitatively, this means that the resonantly excited optical mode is a large susceptibility proxy that applies an amplified effective field that parametrically excites the acoustic mode.

%

\section{Non-linear eigenmode frequencies} \label{NLomegas}
This pulling of the mean state is particularly relevant for the subsequent population dynamics because it changes the natural frequencies of the two eigenmodes through non-linear (NL) frequency shift. 
We have calculated the frequency of the acoustic mode when precessing about the (hypothetically static) mean state. A trivial extension of the dynamical matrix approach \cite{grimsditch_magnetic_2004} yields $\omega_\textrm{acNL}=\gamma_0 \sqrt{H_x(H_x+M_s \langle m_x \rangle)}$. To first order in  $ \langle m_x  (t)\rangle-m_x^\textrm{eq} = \delta m_x  \ll 1$, it is for $H_x=H_\textrm{3MS}$:
\begin{equation} \label{NLomegaAc}
\omega_\textrm{acNL} \approx\omega_\textrm{ac}+\frac12 \gamma_0 M_s \sqrt{\frac {H_j}{H_j+M_s}} \delta m_x \\
\end{equation}
About the mean state, the frequency of the optical mode is $\gamma_0 \sqrt{\mathcal H^2}$ with $\mathcal H^2=H_x^2+H_j M_s + H_x (M_s-2H_j) m_x  + H_j(H_j-2 M_s) (m_x )^2$. At $H_\textrm{3MS}$ it reduces to:
\begin{equation}
\omega_\textrm{opNL} \approx\omega_\textrm{op}-\frac34 \gamma_0 M_s  \sqrt{\frac {H_j}{H_j+M_s}}  \delta m_x  \label{NLomegaOp}
\end{equation}

%
\begin{figure} 
{\includegraphics[width=8 cm]{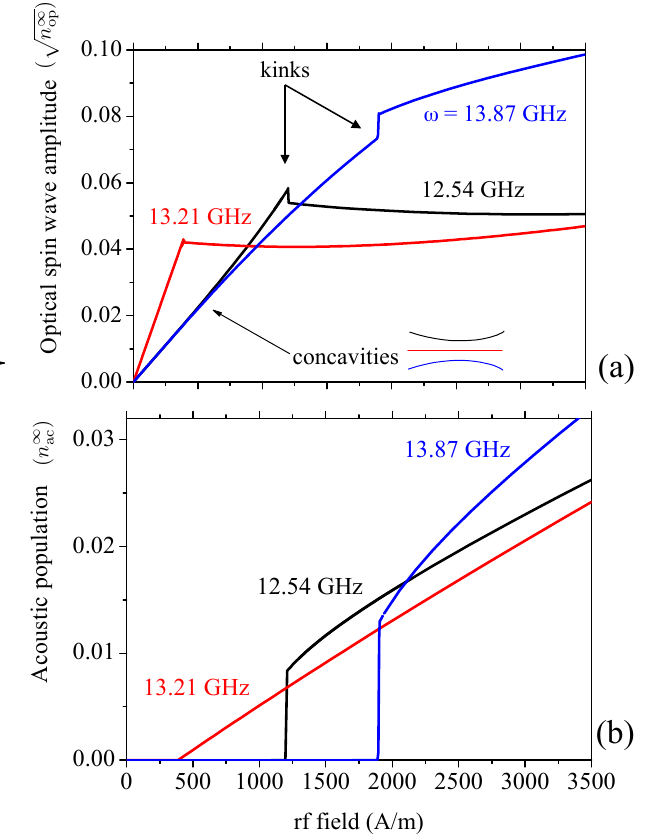}}
{\caption{Final amplitude of the optical mode (a) and final population of the acoustic modes (b) for $H_x=H_{3MS}^\textrm{macrospins}$= 51.3 kA/m (64.42 mT) and for three stimulus frequencies: $\omega_\textrm{rf}=\omega_\textrm{op}$ (red, resonant pumping) and $\omega_\textrm{rf}=\omega_\textrm{op} - 1.1{\Delta \omega_\textrm{op}}$ (black) and $\omega_\textrm{rf}=\omega_\textrm{op} + 1.1{\Delta \omega_\textrm{op}}$ (blue). (a) Final amplitude $\sqrt{n_\textrm{op}^\infty}$ of the optical mode. (b) Final population ${n_\textrm{ac}^\infty}$ of the acoustic mode.}  \label{FinalPopsVersusFreqDetuning}}
\end{figure}

\section{Maximizing the final mode populations by detuning}\label{PopulationMaximization} 
The consequence of frequency detuning on the final populations is displayed in Fig.~\ref{FinalPopsVersusFreqDetuning}. Owing to the Lorentzian nature of the susceptibility of the optical mode at low rf powers, the detuning initially reduces $\sqrt{n_\textrm{op}^\infty}$ by essentially the same amount for $\omega_\textrm{rf} > \omega_\textrm{op}$ and $\omega_\textrm{rf} < \omega_\textrm{op}$ [see Fig.~\ref{FinalPopsVersusFreqDetuning}(a)]. Still below the thresholds but at larger stimuli, a curvature appears indicating that the susceptibility changes when at large levels of optical population. It is a consequence of the negative self non-linear frequency shift of the optical mode, which further amplifies (reduces) the effective detuning when $\omega_\textrm{op} < \omega_\textrm{rf} $ (when $ \omega_\textrm{op} > \omega_\textrm{rf}$), thereby decreasing (increasing) the marginal susceptibility as $n_\textrm{op}$ increases. 

Above the threshold, the dependence of the populations (Fig.~\ref{FinalPopsVersusFreqDetuning}) exhibits kinks and complex curvatures: it is far more complicated that one would infer from just the $\omega_\textrm{rf}$-dependence of the threshold displayed in Fig.~1(b) of the main document. In a counterintuitive manner, the two mode populations can both reach final levels that are \textit{larger} when working off-resonance. 

This counterintuitive result arises from the non-linear frequency shifts of the eigenmodes.
The NLFS's have a practical consequence: when resonantly pumping in perfectly matching condition using $\omega_\textrm{rf}=\omega_\textrm{op}= 2\omega_\textrm{ac}$, the NLFS pushes immediately the system out of the (linear) resonance condition, yielding $\omega_\textrm{rf} \neq \omega_\textrm{opNL}\neq  2\omega_\textrm{acNL}$. This inevitably limits the efficiency of the direct pumping of the optical mode, reducing in fine the ability of the rf field to inject power into the modes. The formerly observed (see the beginning of the shaded areas of Fig.~\ref{TimeResolvedPopulations1kAperMeter}) small reduction of the stored energy and of the optical population seen just after $t=2/\Delta\omega_\textrm{op}\approx 1 $ ns in nominally resonant pumping originates from this optical-population induced detuning.

We can now understand the surprising discontinuity ("kink") observed for $\omega_\textrm{rf} > \omega_\textrm{op}$ (Fig.~\ref{FinalPopsVersusFreqDetuning}(a), blue curve). As already said, the negative curvature of $n_\textrm{op}^\infty (H_x^{rf})$ in the sub-threshold regime is due to the negative self-NLFS ($\frac{d\, \omega_\textrm{op} } {d\,n_\textrm{op}} <0$), which gradually increases (worsens) the detuning. However when the 3MS threshold is reached, the sudden increase of the acoustic population activates the $\frac{d\, \omega_\textrm{op} } {d\,n_\textrm{ac}} >0$ frequency shift and restores a condition closer to resonance for the optical mode, which results in the counterintuitive positive kink seen in Fig.~\ref{FinalPopsVersusFreqDetuning}(a) (blue curve) at the threshold. Since the optical population reaches a level higher than when resonantly pumping in perfectly matching condition, the acoustic population also does. The same rational can be applied to the negative detuning case [Fig.~\ref{FinalPopsVersusFreqDetuning}, black curves]. For $\omega_\textrm{rf} < \omega_\textrm{op}$ a positive curvature is seen for the optical population below the threshold as a result of negative self-NLFS reducing the detuning, followed by a negative kink the optical population at the threshold resulting from the positive cross-NLFS.

Pre-detuning on purpose the applied frequency can pre-compensate for the NLFS-induced detuning, so that the obtained population can effectively reach higher levels provided the threshold is overcome (see Fig.~\ref{FinalPopsVersusFreqDetuning}). The stimulus frequency $\omega_\textrm{max}$ that maximizes the energy injected in the eigenmode system should satisfy the transcendantal equation $\omega_\textrm{opNL}=\omega_\textrm{max}$ in the final state. To first order in $\delta m_\textrm{eq}$, this can be re-written as:
 \begin{multline}
\omega_\textrm{max}- \omega_\textrm{op} =-\frac34 \gamma_0 M_s  \sqrt{\frac {H_j}{H_j+M_s}} \\ \times \left( \underbrace{\frac {d\, {\langle m_x }\rangle}  {d\, {n_\textrm{op}}} }_{\approx 0.93}n_\textrm{op}^\infty + \underbrace{\frac {d\, {\langle m_x }\rangle}  {d\, {n_\textrm{ac}}}}_{\approx -1.82} n_\textrm{ac}^\infty \right)
 \end{multline}
When substantially above the 3MS threshold, we have $n_\textrm{ac}^\infty > n_\textrm{op}^\infty$ such that the term in parenthesis is negative and the stimulus frequency $\omega_\textrm{max}$ that maximizes the inductive injection of energy into the eigenspectrum is larger than $\omega_\textrm{op}$. This could of course have been anticipated from a comparison of the curves in Fig.~\ref{FinalPopsVersusFreqDetuning}. 

\section{Phase diagrams}\label{PhaseDiagrams} 
%
\begin{figure*} 
{\includegraphics[width=18 cm]{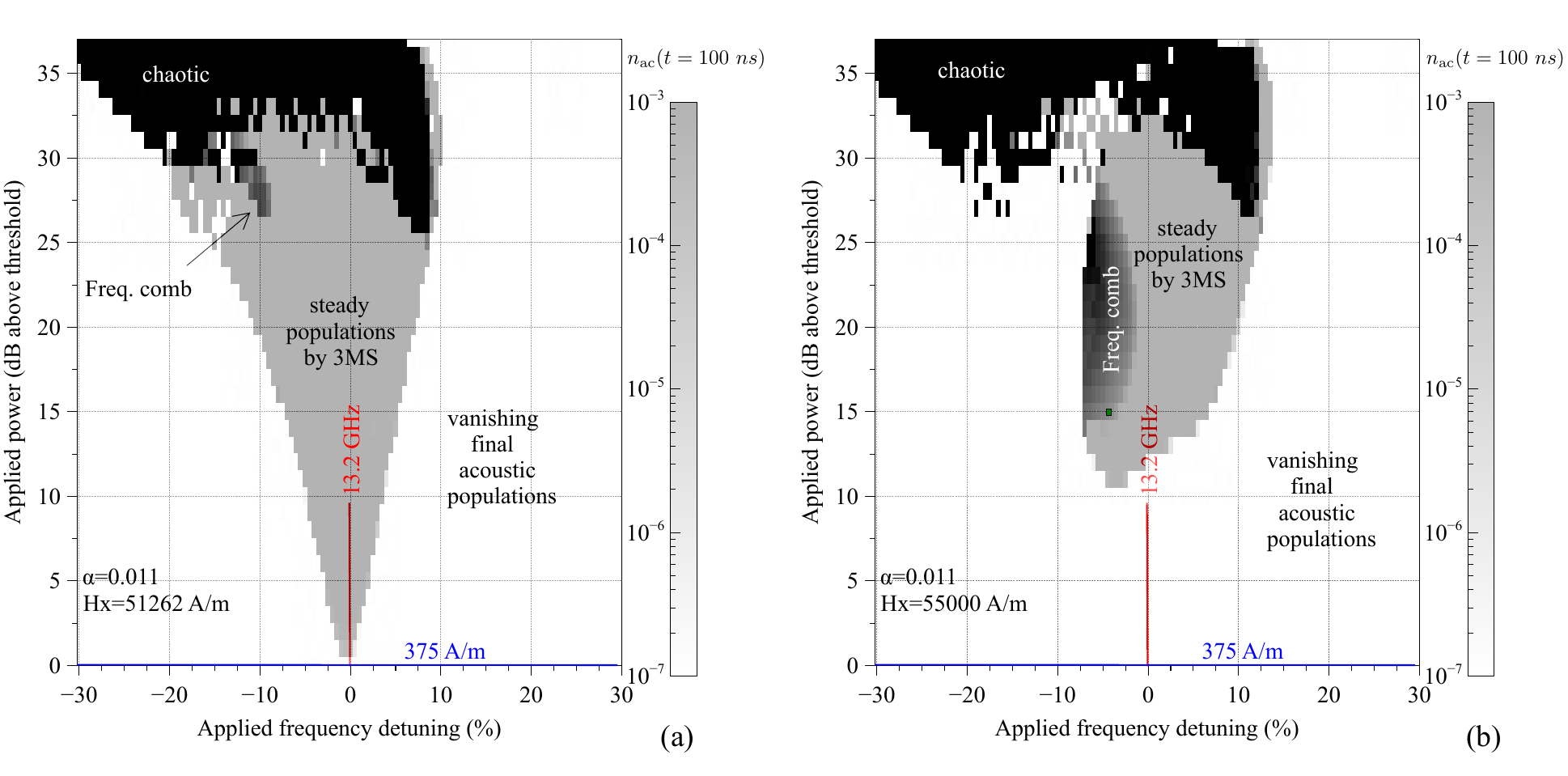}}
\caption{Phase diagrams of the dynamics after 100 ns of application of the stimulus, starting from the thermal seed state. In the white region, the final acoustic population vanishes. The black region corresponds to the absence of a steady state and the absence of population periodicity. The gray area corresponds to regions where there are periodic oscillations of the spin wave populations leading to a frequency comb. (a) At the field $H_x=H_\mathrm{3MS}= 51263$ A/m (64.4 mT) where $\omega_\mathrm{op} = 2 \omega_\mathrm{ac}$ and leads to  $\omega_\mathrm{op} / (2 \pi)=13.2$ GHz. (b) At the field $H_x=$ 55000 A/m (69.1 mT) $> H_\mathrm{3MS}$. The green rectangle corresponds to the calculation of Fig. 4.} \label{FigurePhaseDiagrams}
\end{figure*}

Two categories of behaviors were shown so far. Fig. 2 illustrated conditions for which finite steady state populations are achieved when 3MS is active and NLFS is moderate. This was obtained for $H_x=H_\mathrm{3MS}$ and $\omega_\textrm{rf}=\omega_\mathrm{op} = 2 \omega_\mathrm{ac}$. In contrast,  Fig. 4 showed a situation leading to the periodic growth and decay of the two mode populations. This was obtained for a different field and a detuned frequency, i.e. $H_x > H_\mathrm{3MS}$ and $\omega_\textrm{rf} < \omega_\mathrm{op} < 2 \omega_\mathrm{ac}$. The two categories of dynamical regime can be prepared by proper choices of the applied frequency, applied power and applied dc field.

This is illustrated in Fig.~\ref{FigurePhaseDiagrams} which shows the phase diagrams for the two applied fields formerly used of Fig.~2 and Fig.~4, both when starting from the thermal seed condition (Eq.~\ref{ThermalFluctuationAmplitude}). The white regions at low rf field are below the threshold for 3MS, so that the final population of the acoustic mode vanishes. The grey areas mark the regions where the final population of the acoustic mode converges to a non-vanishing stable value. The black regions labeled "chaotic" are the ones when the populations of the two modes neither stabilize to a fixed point not oscillate in a periodic manner. Finally, the dark grey regions labeled "freq. comb" indicate the self-induced Floquet regions where the populations undergo cyclic growth and decay as exemplified in Fig. 4. These regions forms a pocket in the $\{f_\textrm{rf}, H_x^\textrm{rf}\}$ space. This pocket enlarges for low damping values and gradually shrinks when increasing the damping to 0.018, where the pocket disappears. The abrupt boundary at negative frequency detuning probably indicates the existence of several dynamical states that may be prepared by varying the initial conditions away from the regular thermal seed \textcolor{black}{(see next section)}.

\begin{figure} 
\begin{centering}
\includegraphics[width=8.5 cm]{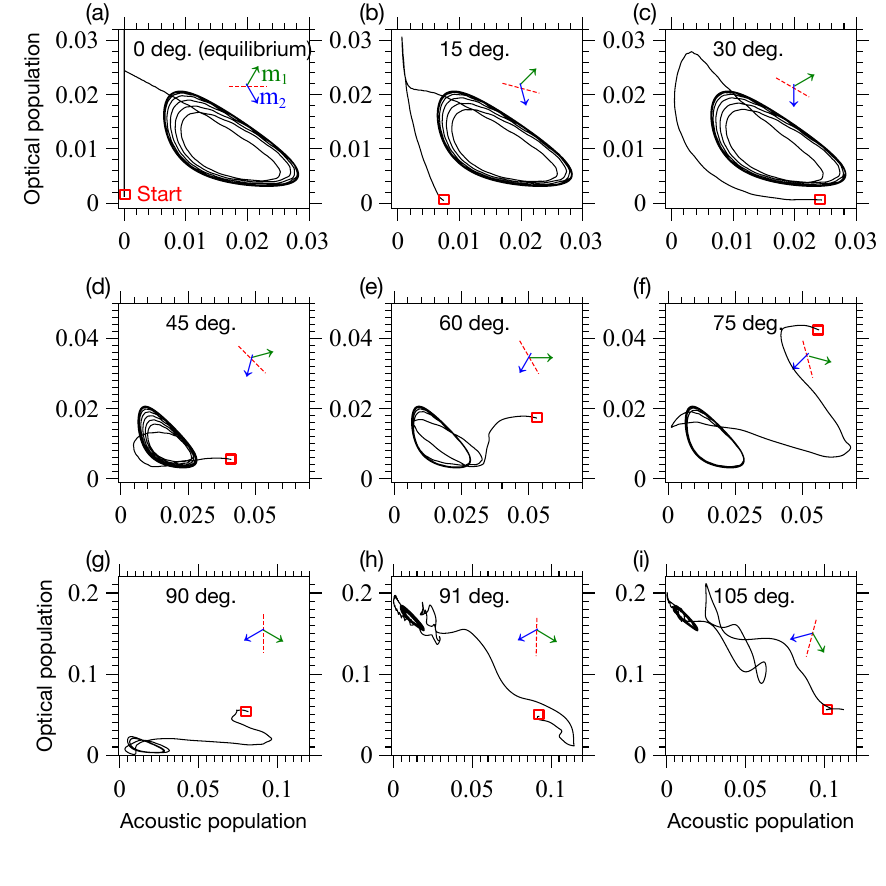}
\end{centering}
\caption{Population dynamics in the conditions of Fig. 4 (i.e. $H_x$=55 kA/m, $H_x^\textrm{rf}=2.2$ kA/m and $\omega/(2 \pi)=12.68$ GHz) for initial conditions at equilibrium or far from equilibrium. (a): When the initial magnetizations are in the scissor state with thermal fluctuations at $T=300$ K according to Eq.~\ref{ThermalFluctuationAmplitude}.  (b-i): When the initial magnetizations are rotated in the sample plane by 15, 30, 45, 60, 75, 90, 91 and 105 degrees. The two last conditions lead to a different limit cycle for the spin wave populations.}
\label{InfluenceOfInitialState}
\end{figure}
\textcolor{black}{
\section{Influence of the initial magnetic state on the self-induced Floquet state}
\label{InfluenceInitialState}
All the simulations shown so far were conducted assuming initial magnetizations in the scissor state with thermal fluctuations at $T=300$ K applied on the sole acoustic mode according to Eq.~\ref{ThermalFluctuationAmplitude}. Since the dynamics is non-linear, it is important to study the influence of the initial state on the resulting dynamics, especially in the case of the cyclic growth and decay of the mode populations leading to the self-induced Floquet state. The phase space of the initial state is 4-dimensional and can be indexed either by the two orientations of each magnetization, or by the two mode populations and their phases. Only part of this phase space may be accessible in experiments, notably by designing sequences of applied fields.}

\textcolor{black}{We first investigated the starting from scissors states of various openings about the applied field, i.e. $ \mathbf{m}_1(t=0)=(\cos \theta, \sin \theta, 0)$ and $ \mathbf{m}_2(t=0)=(\cos \theta, - \sin \theta, 0)$, with variable $\theta \in [0, 2 \pi]$. This is equivalent to starting from a variable (and very large) optical population and a vanishing acoustic population. Such a situation is not achievable by thermal fluctuations. However, this would be feasible experimentally if the strength of the applied "dc" field $H_x$ would be abruptly changed at $t=0$. All these initial states were found to always lead to the same dynamics (not shown), with populations finally orbiting around the Floquet limit cycle already evidenced in Fig. 4.}

\textcolor{black}{We then investigated the starting from scissors states that are rotated in the sample plane out of the equilibrium opening $\theta_\textrm{eq}$, i.e. by choosing $ \mathbf{m}_1(t=0)=(\cos (\theta_\textrm{eq}+\theta), \sin (\theta_\textrm{eq}+\theta), 0)$ and $ \mathbf{m}_2(t=0)=(\cos (\theta-\theta_\textrm{eq}), \sin (\theta - \theta_\textrm{eq}), 0)$, with variable $\theta \in [0, 2 \pi]$. This changes the two initial populations: the change of acoustic population is intuitive since the acoustic mode is a rigid rocking of the scissors state. The change of the optical population can be understood by noticing that a rotation of $\theta = \pi$ is equivalent to changing the scissoring angle, which is a change of the optical population. Such a situation is also not achievable by thermal fluctuations could be prepared if the applied "dc" field was abruptly rotated at $t=0$ and the stimulus was turned on at that same time. We have found that if the rotation is less than 90 degrees, the populations still converge to the same limit cycle. For larger rotations (panels (h) and (i) of Fig.~\ref{InfluenceOfInitialState}), the final state depends sensitively on the initial state and can either be the same limit cycle, or another limit cycle with a much larger optical population but still a cyclic growth and decay of the two populations.}

\begin{figure} 
\begin{centering}
\includegraphics[width=8.5 cm]{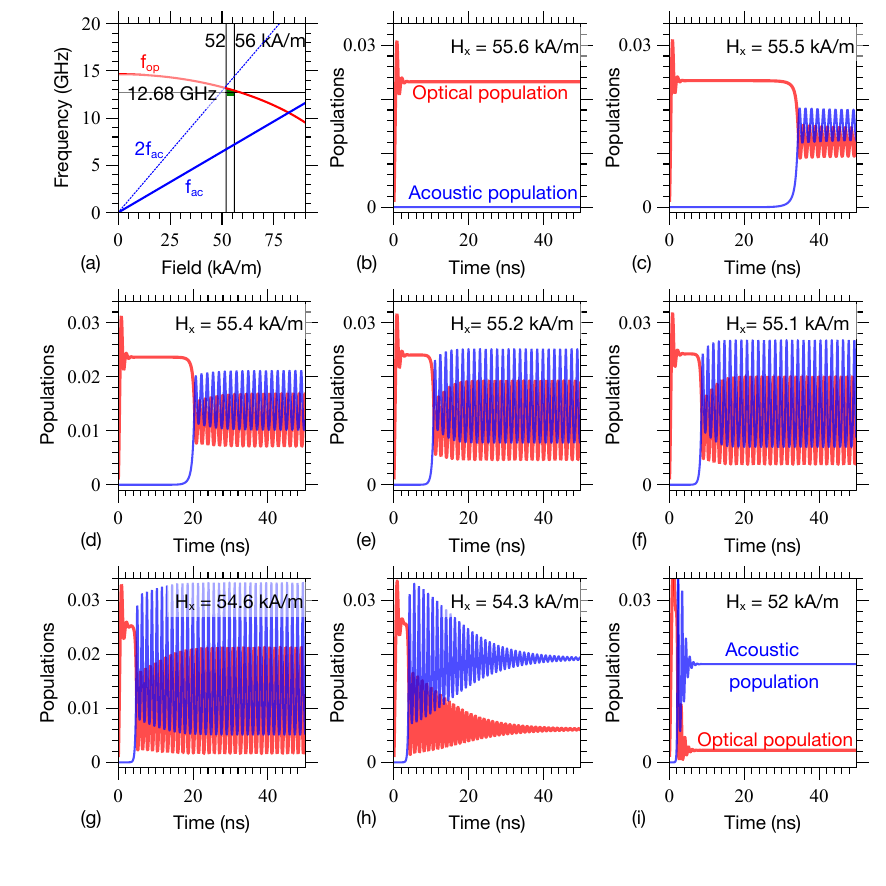}
\end{centering}
\caption{Population dynamics for variable $H_x$=52-56 kA/m at fixed stimulus amplitude $H_x^\textrm{rf}=2.2$ kA/m and fixed applied frequency $\omega_\textrm{rf}/(2 \pi)=12.68$ GHz. The initial sate is always taken according to Eq.~\ref{ThermalFluctuationAmplitude}.  (a) Mode frequencies versus field. The green bar is at the applied frequency and spans over the range of investigated applied fields. (b-i): population dynamics for fields indicated in each panel. }
\label{InfluenceOHxField}

\end{figure}

\textcolor{black}{\section{Effect of detuning on the self-induced Floquet states}
\label{Detuning}
Since the dynamics in the self-induced Floquet state is the result of the interplay between the initial frequency detuning and the crossed- and self-non-linear frequency shifts, we can tune this dynamics by changing the order between the applied frequency $\omega_\textrm{rf}$, the frequency of the optical mode $\omega_\textrm{op}$ in the linear regime and that of the pair of acoustic modes $2 \omega_\textrm{ac}$. These three frequencies can be ordered in 6 different manners by spanning the in-plane field $H_x$ symmetrically about $H_\textrm{3MS}=51.3$ kA/m (64.4 mT) and by spanning the applied frequency symmetrically about $\omega_\textrm{op}(H_\textrm{3MS})=13.2$ GHz.}

\textcolor{black}{We have investigated these 6 possibilities for a stimulus of amplitude $H_x^\textrm{rf}=2200$ A/m. Among the 6 detuning possibilities, only the configuration obeying 
\begin{equation} \omega_\textrm{rf} < \omega_\textrm{op} < 2 \omega_\textrm{ac} \end{equation}
leads to the self-induced Floquet states associated to the cyclic growth and decay of the two mode populations and the resulting beating of the period averaged mean state. Fig.~\ref{InfluenceOHxField} reports the population dynamics in such a situation for several representative dc applied fields $H_x$ in the 52-56 kA/m interval,  i.e. for $H_x>H_\textrm{3MS}$ [The investigated region is highlighted as the green color in Fig.~\ref{InfluenceOHxField}(a)]. }

\textcolor{black}{At large $H_x$ [Fig.~\ref{InfluenceOHxField}(b)], the optical mode is almost resonant and can be therefore pumped at a high initial rate. However at this same large field the acoustic mode is strongly detuned (i.e. $\omega_\textrm{ac} - \omega_\textrm{rf}/2 \gg \Delta\omega_\textrm{ac}$), such that the effective field supplied by the optical mode (Eq.~\ref{EffectiveFieldFromOpticalMode}) does not induce frequent 3MS events and the acoustic population would vanish in the absence of NLFS. The sole NLFS to be playing a role are thus the ones resulting from $n_\mathrm{op}$. Since $\frac{\partial \omega_\mathrm{ac}}{\partial n_\mathrm{op}} >0$, the acoustic mode gets even more detuned after the initial dynamics and therefore stays its population stays at its vanishing level forever.}

\textcolor{black}{In the opposite case at low fields [Fig.~\ref{InfluenceOHxField}(h,i)], it is the optical mode that is strongly detuned and is therefore initially pumped a low rate. Meanwhile the acoustic mode is now close to resonance, so that almost each created optical mode can immediately undergo 3MS, which raises the population  of the acoustic mode to a high level. After this initial phase, the main NLFS at play are thus solely the two ones resulting from $n_\mathrm{ac}$.
Since $\frac{\partial \omega_\mathrm{op}}{\partial n_\mathrm{ac}} >0$ and $\frac{\partial \omega_\mathrm{ac}}{\partial n_\mathrm{ac}} <0$, the optical mode gets even more detuned while the acoustic mode gets even closer to resonance, which reinforces the initial evolution of the populations. As a result,  the optical population is bound to stabilize at a low level and the acoustic one at a large level [Fig.~\ref{InfluenceOHxField}(h,i)].} 

\textcolor{black}{The interesting situation is the one occurring at intermediate fields [Fig.~\ref{InfluenceOHxField}(c-g)], when the two initial detunings are both moderate so that the two modes can be populated in the first place, such that the 4 NLFSs can then induce a substantial feedback leading to a complex dynamics.
The cyclic growth and decay of the two mode populations requires two conditions that can be expressed in a semi-quantitative manner. The first one is to populate the optical mode in the initial dynamics. Qualitatively, this simply reads $$ \omega_\textrm{op} - \Delta \omega_\textrm{op} < \omega_\textrm{rf}. $$
The complementary condition $ \omega_\textrm{rf} <  \omega_\textrm{op} + \Delta \omega_\textrm{op}$ is inappropriate to induce the complex dynamics.
Getting a limit cycle of the two mode populations requires additionally to have a negative feedback from the populations onto the growth rate of the optical mode. There need to be a time at which the optical mode NLFS has become large enough to completely detune the optical mode from the applied rf drive, so that the optical population suddenly collapses. A necessary condition for this NLFS-induced detuning of the optical mode is that at some given time $t$ we have:
$$ \omega_\textrm{rf} <  \omega_\textrm{op} - \Delta \omega_\textrm{op} + \frac{\partial \omega_\textrm{op}} {\partial n_\textrm{ac}} n_\textrm{ac}(t)+ \frac{\partial \omega_\textrm{op}} {\partial n_\textrm{op}} n_\textrm{op}(t)$$
which leads to the condition: 
\begin{equation} 
\label{ExistenceFloquet}
\exists t,~ \frac{n_\textrm{ac}(t)}{n_\textrm{op}(t)} > - \left({\frac{\partial \omega_\textrm{op}} {\partial n_\textrm{op}}} / {  \frac{\partial \omega_\textrm{op}} {\partial n_\textrm{ac}} }  \right) \\
\end{equation} where the RHS term is strictly positive thanks to Eq.~\ref{NLFS}.
Meanwhile, the acoustic mode must stay close to the resonance condition for 3MS to happens, which leads to the additional semi-quantitative condition: 
\begin{equation} 
\label{ExistenceFloquet2}
\forall t,  ~- \left({\frac{\partial \omega_\textrm{ac}} {\partial n_\textrm{op}}} / {  \frac{\partial \omega_\textrm{ac}} {\partial n_\textrm{ac}} }  \right) > \frac{n_\textrm{ac}(t)}{n_\textrm{op}(t)}  
\end{equation} 
The threshold of Eq.~\ref{ExistenceFloquet} is never reached at high field while Eq.~\ref{ExistenceFloquet2} is not satisfied at low field.
Qualitatively, if the feedback condition of Eq.~\ref{ExistenceFloquet} is first met at some time $t_0$ the optical population immediately starts to decay at the rate $\Delta\omega_\textrm{op}$. The 3MS ensured by Eq.~\ref{ExistenceFloquet2} postpones the immediate decay of the acoustic population, that would otherwise decay at approximately the same natural rate since $\Delta\omega_\textrm{ap} \approx \Delta\omega_\textrm{op}$. The ratio $\frac{n_\textrm{ac}(t)}{n_\textrm{op}(t)}$ thus further increases at $t>t_0$. This leads to a collapse of $n_\textrm{op}(t)$ until we get back in resonance condition when the populations have sufficiently dropped. }

%
\begin{figure} 
\includegraphics[width=8 cm]{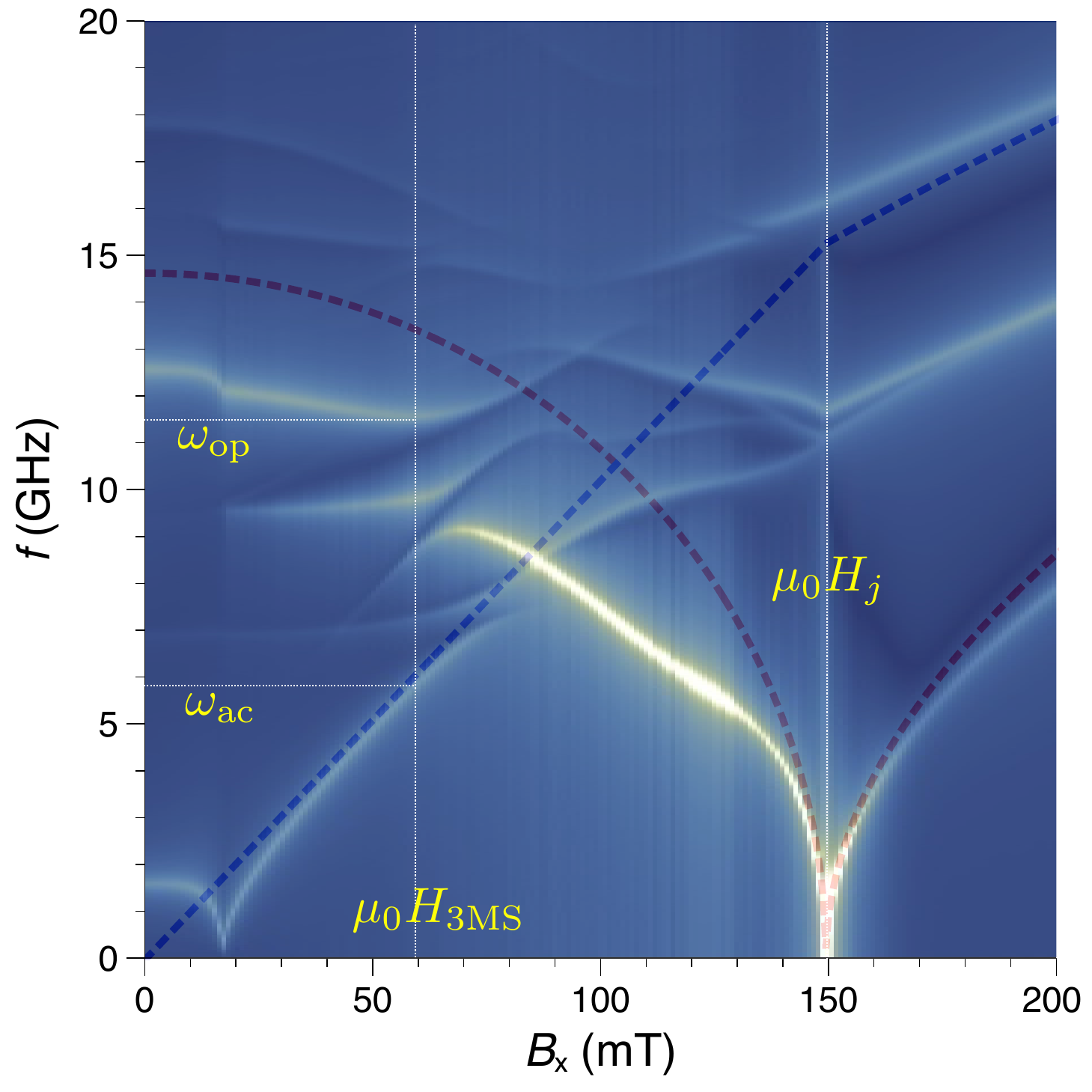}
\caption{Comparison of the eigenmodes of two SAFs of saturation field $\mu_0 H_j$=150 mT, either in the two-macrospin model (dotted lines) or in the micromagnetic description (color map). The dotted blue (red) line is the acoustic (optical) mode of the two-macrospin model throughout the paper. The color map is the superposition of the susceptibility maps of the in-plane (i.e. $m_{x1}$) and out-of-plane (i.e. $m_{z1}$) componenent of a CoFeB SAF nanodot of 100 nm diameter, $2\times2~\textrm{nm}$ thickness and interlayer exchange coupling of -0.08 mJ/m$^2$. Each susceptibility map is the power spectrum calculated from the impulse response to a uniform field along $z$. }
\label{MicromagneticModes}

\end{figure}

%
\begin{figure} 
\includegraphics[width=8 cm]{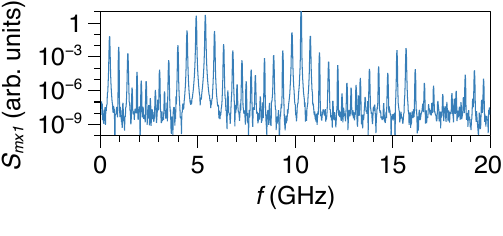}
\caption{Micromagnetic calculation of the spectrum of the x-component of the magnetization of the 2 nm-thick top layer of a SAF nanodot submitted to an in-plane dc field of 49.3 kA/m (62 mT) and an rf field of 3.98 kA/m (5 mT). The dot is a disk of 100 nm diameter and interlayer exchange coupling of -0.08 mJ/m$^2$. }
\label{mumax}
\end{figure}

\section{Comparison to micromagnetics}\label{micromagnetics} 

 Our model relies on the crude two-macrospin approximation. In this last section, we use micromagnetic simulations to show that a frequency comb arising from the cyclic growth and decay of the two main spin wave populations can also be obtained in a realistic SAF nanodot. We choose a circular system with 100~nm diameter made of a CoFeB (2 nm) / spacer / CoFeB (2 nm) SAF. The micromagnetic simulations are done using the simulation solftware mumax3 \cite{vansteenkiste_design_2014} \textcolor{black}{with $64\times64\times2$ cells. The magnetization is 1.25 MA/m, the exchange stiffness is 16 pJ/m and the damping is 0.011.} An interlayer exchange coupling of -0.08 mJ/m$^2$ was chosen so as to lead a saturation fields of 150 mT, similar to that of the two-macrospin model used in the main part of the manuscript. Note that this value of $J$ is a moderate interlayer exchange coupling, as encountered for instance when Cu is used as a spacer \cite{stiles_interlayer_1999}. 
 
\textcolor{black}{Fig.~\ref{MicromagneticModes} compares the eigenexcitations of the micromagnetic nanodot with that of the two-macrospin system used throughout the paper. The additional degrees of freedom inherent to the micromagnetic description complicate the eigenexcitation spectra, which exhibit many more modes as well as anticrossings evidencing their hybridizations. However, there is still one high susceptibility mode that softens at $\mu_0 H_x$=150 mT and that possesses an overall dispersion $\omega(H_x)$ that bears a large similarity with the optical mode of the two-macrospin system. Besides, the micromagnetic eigenexcitations also comprise a mode whose frequency is rather linear with the field and ressembles the acoustic one of the two-macrospin system. Interestingly, we can also define an applied field $H_\textrm{3MS}$ for which the matching condition $\omega_\textrm{op} \approx 2 \omega_\textrm{ac}$ is realized within the micromagnetic description (dashed lines in Fig.~\ref{MicromagneticModes}). It is $\mu_0 H_\textrm{3MS}=58$ mT and it leads to $\omega_\textrm{op}/(2 \pi)=11.4$ GHz, to be compared to the 62 mT and 13.2 GHz formerly used in the two-macrospin model (see the comparison in Fig.~\ref{MicromagneticModes}).}

\textcolor{black}{Following the rational of the main paper, we can increase the applied field by circa 5 mT and decrement the stimulus frequency by circa one linewidth to keep finite susceptibilities while achieving the criterion $\omega_\textrm{rf} < \omega_\textrm{op} < 2 \omega_\textrm{ac}$ that was formerly identified to favor the formation of a self-induced Floquet state.}
The micromagnetic response of the SAF nanodot is evaluated in Fig.~\ref{mumax}. The applied dc field is 49.3 kA/m (62 mT) and prepares a scissors state. An rf field of 3.98 A/m (5 mT) at 10.3 GHz is applied. The power spectrum of the spatial average of $m_{x1}(t)$ in the steady state is displayed is Fig.~\ref{mumax}. This spectrum calculation within the micromagnetic framework bears a large similarity with that obtained in the two-macrospin approximation earlier in Fig. 4(e) of the main document. This demonstrates that the main effect --the obtention of self-induced Floquet states via three-wave processes in synthetic antiferromagnets-- may arise in realistic systems, if following the rational inspired from the two-macrospin model.

\end{document}